\documentclass{article}

    \usepackage[preprint]{neurips_2024}

\usepackage[utf8]{inputenc} 
\usepackage[T1]{fontenc}    
\usepackage{hyperref}       
\usepackage{url}            
\usepackage{booktabs}       
\usepackage{amsfonts}       
\usepackage{amsmath}
\usepackage{nicefrac}       
\usepackage{microtype}      
\usepackage{xcolor}         
\usepackage{multirow}
\usepackage{booktabs,caption}
\usepackage{threeparttable}
\usepackage{graphicx}
\usepackage{placeins}

\makeatletter
\AtBeginDocument{%
  \expandafter\renewcommand\expandafter\section\expandafter
    {\expandafter\@fb@secFB\section}%
  \newcommand\@fb@secFB{\FloatBarrier
    \gdef\@fb@afterHHook{\@fb@topbarrier \gdef\@fb@afterHHook{}}}%
  \g@addto@macro\@afterheading{\@fb@afterHHook}%
  \gdef\@fb@afterHHook{}%
}
\makeatother

\makeatletter
\AtBeginDocument{%
  \expandafter\renewcommand\expandafter\subsection\expandafter
    {\expandafter\@fb@secFB\subsection}%
}
\makeatother

\title{How Students Use AI Feedback Matters: 
Experimental Evidence on Physics Achievement and Autonomy}

\author{%
  Xusheng Dai \\
  School of Education\\
  Tsinghua University\\
  Beijing, China \\
  \texttt{dxs23@mails.tsinghua.edu.cn} \\
  \And
  Zhaochun Wen \\
  School of Education\\
  Tsinghua University\\
  Beijing, China \\
  \texttt{wzc23@mails.tsinghua.edu.cn} \\
  \And
  Jianxiao Jiang \\
  School of Education\\
  Tsinghua University\\
  Beijing, China \\
  \texttt{jjx23@mails.tsinghua.edu.cn} \\
  \And
  Huiqin Liu \\
  School of Education\\
  Tsinghua University\\
  Beijing, China \\
  \texttt{liuhq@tsinghua.edu.cn} \\
  \And
  Yu Zhang\thanks{Corresponding to Dr. Yu Zhang at zhangyu2011@tsinghua.edu.cn. Corresponding author address
417 Wennan Building, Tsinghua University, Beijing, China, 100084.} \\
  School of Education\\
  Tsinghua University\\
  Beijing, China \\
  \texttt{zhangyu2011@tsinghua.edu.cn} \\
}

\begin{document}

\maketitle

\begin{abstract}
  Despite the precision and adaptiveness of generative AI (GAI)-powered feedback provided to students, existing practice and literature might ignore how usage patterns impact student learning. This study examines the heterogeneous effects of GAI-powered personalized feedback on high school students' physics achievement and autonomy through two randomized controlled trials, with a major focus on usage patterns. Each experiment lasted for five weeks, involving a total of 387 students. Experiment 1 (n = 121) assessed compulsory usage of the personalized recommendation system, revealing that low-achieving students significantly improved academic performance (d = 0.673, p < 0.05) when receiving AI-generated heuristic solution hints, whereas medium-achieving students' performance declined (d = -0.539, p < 0.05) with conventional answers provided by workbook. Notably, high-achieving students experienced a significant decline in self-regulated learning (d = -0.477, p < 0.05) without any significant gains in achievement. Experiment 2 (n = 266) investigated the usage pattern of autonomous on-demand help, demonstrating that fully learner-controlled AI feedback significantly enhanced academic performance for high-achieving students (d = 0.378, p < 0.05) without negatively impacting their autonomy. However, autonomy notably declined among lower achievers exposed to on-demand AI interventions (d = -0.383, p < 0.05), particularly in the technical-psychological dimension (d = -0.549, p < 0.05), which has a large overlap with self-regulation. These findings underscore the importance of usage patterns when applying GAI-powered personalized feedback to students.
\end{abstract}

\section{Introduction}

Driven by the rapid development of generative artificial intelligence (GAI), numerous technological innovations have emerged to enrich K–12 education, particularly in STEM domains where robust logical reasoning and sophisticated pedagogical design are essential (García-Martínez et al., 2023). Although prior research on technology‑enhanced STEM learning has delivered significant advances in personalization and automation (Smith et al., 2023), it has tended to overlook the usage patterns and the development of students’ ability of learning. For example, in one semester‑long intervention offering individualized support, participants’ end‑of‑term grades declined, suggesting that an exclusive focus on tailored content delivery may inadvertently neglect the cultivation of essential learning skills (Bastani et al., 2024).

The usage patterns reflect pedagogy and rules of student learning in nature, while most studies only focus on the precision and adaptiveness of contents (Maier \& Klotz, 2022). The dominant idea of personalized feedback mainly focuses on algorithm and pre-defined usage pattern, but neglect students’ characteristics and their needs (Zheng \& Han, 2024). As illustrated in \textit{Xue Ji}, chapter of \textit{Liji} (The Book of Rites), “an effective response to inquiry resembles the striking of a bell: a light strike produces a soft sound, while a heavy strike yields a resonant tone.” (Legge, 1967) This metaphor underscores the need for responsive teaching that aligns with learners’ cognitive readiness and curiosity—something many algorithmic systems fail to account for. 

Regarding the learning outcome, both learning ability and academic achievement are crucial, representing related but distinct constructs. Academic achievement denotes the level of mastery demonstrated by assessments, whereas learning ability refers to the combination of cognitive, metacognitive, and motivational capacities that enable individuals to acquire and apply knowledge effectively across various contexts (UNESCO, 2022). Within this framework, autonomy plays a pivotal role. Holec (1981) defines it as "the ability to take charge of one's own learning" (p. 3). When learners perceive autonomy in their educational environment, they are more likely to exhibit intrinsic motivation, leading to deeper engagement and improved learning outcomes (Deci \& Ryan, 2000). Autonomous learners tend to actively engage in self-regulated learning (SRL) (Oguguo et al., 2023), a process widely recognized as a key driver of academic success and lifelong development (Boyraz et al., 2016). These contemporary understanding of autonomy aligns with the philosophical perspective wherein moral agency arises from individuals acting according to principles they have rationally chosen for themselves (Kant, 1992). 

Despite these theoretical insights, empirical research in AI in education (AIED) has predominantly targeted achievement outcomes, with relatively few studies examining effects on autonomy. Some scholars have cautioned that extensive reliance on AI tools may foster technical dependency and undermine independent thinking and metacognitive monitoring (Hyde et al, 2024), yet rigorous, long‑term investigations of these potential risks remain scarce. Moreover, existing studies often adopt a limited practical approach, primarily facilitating direct interactions between students and ChatGPT without systematically exploring well designed learning systems underpinned by pedagogical frameworks and empowered by GAI (Deng et al., 2025).

Building on the usage pattern categorization proposed by Du et al (2023) and Lai (2015), two primary AI‑powered pedagogies in personalized learning were distinguished: (a) adaptive learning or personalized recommendation—wherein the system proactively creates learning pathways or resources based on learner data; and (b) adaptable learning or on‑demand help—wherein learners autonomously request just‑in‑time explanatory or problem‑solving assistance. Apart from these patterns, variations in subject, educational level and intervention duration have further impeded cross‑study comparability, resulting in limited external validity for practitioners.  

Beyond the imperative to boost immediate academic performance, preserving learner autonomy is vital priority for the next generation of AI‑enhanced learning. To address these gaps, the present study examines the effect on both achievement and autonomy. In a naturalistic school context, two random experiments were conducted sequentially to evaluate the impacts of personalized recommendation and on‑demand help on students with varying baseline achievement levels. By integrating measures of autonomy alongside traditional performance metrics, this research aims to offer a more comprehensive understanding of how GAI‑driven scaffoldings can support sustainable STEM learning.

\section{Literature Review}

\subsection{Educational personalized recommendation}

Educational Personalized Recommendation (EPR) refers to a service that recommends educational content or activities to educators or learners based on algorithms. The idea of EPR originally derived from commercial recommendation systems—which recommends products to consumers based on previously collected customer preference and behavior data. By doing so, such system can reduce decision-making cost, trigger impulse purchasing, and thus boost sales (Nguyen \& Tran, 2024). EPR borrows this logic by using instructional preferences and diagnostic data on learning conditions to recommend content or activities that are considered most beneficial for learners' development (Zhang \& Wang, 2022).

The advent of GAI offers potential breakthroughs in applying recommender systems in educational practice. Retrieval-augmented generation workflows based on pre-trained large language models significantly reduce the need for algorithms and extensive data. Functionalities that previously required absorbing thousands of data points can now be effectively achieved with just a few prompts. Influenced by pro-AI policy and market dynamics, numerous learning apps, smart devices, and intelligent tutoring classrooms integrating multiple GAI-powered recommendation functionalities have emerged rapidly. 

Regarding the implementation strategies of EPR, Deschênes (2020) suggest that most EPR systems focus primarily on recommending academic content such as courses, materials, and exercises. Deschênes argue that EPR systems can maximize knowledge acquisition and even support for agency, since such systems can reduce the time learners spend searching and help them more efficiently access the content most relevant to their needs. Recent empirical studies have increasingly demonstrated the educational value EPR across diverse subjects and learner populations. Sancenon et al. (2022) conducted a randomized controlled trial with 43 Singaporean primary school students, implementing an adaptive recommendation system that customized worksheets based on learning analytics, and found that personalized content delivery significantly improved academic performance. Similarly, Hsu et al. (2013) investigated the effects of mobile-based personalized reading recommendations among senior high school EFL students, assigning participants into two experimental groups (with personal or shared annotations) and a control group, and reported that both experimental groups outperformed the control group in reading comprehension. In a flipped classroom context, Huang et al. (2023) applied AI-enabled personalized video recommendations for 43 college students studying systems programming and compared their outcomes with 59 control participants; results indicated that personalized recommendations significantly enhanced engagement and learning performance among students with moderate initial motivation. Extending the application to mobile learning environments, Drissi et al. (2024) developed GAMOLEAF, a gamified mobile learning framework for teaching Java data structures during COVID-19, and showed through a three-group comparison (N=90) that the integration of both gamification and personalized recommendation features led to significantly higher learning achievement and motivation compared to either intervention alone. Additionally, Dai et al. (2023) explored the role of adaptive testing technology in personalized recommendations for junior middle school students, revealing that recommending slightly more challenging content based on learner ability profiles significantly enhanced students' scores and learning growth compared to random recommendation. Despite these findings, there remains a notable gap of empirical research addressing the potential efficacy of EPR in high school STEM education.  Consequently, randomized controlled trials are imperative to assess and establish the effectiveness of EPR in enhancing STEM learning at the high school level.

Regarding non-academic outcomes, a few studies have examined the effects on psychology, cognition, and behavior. Hsu et al. (2013) found that recommending reading materials aligned with students' proficiency significantly reduced cognitive load. Drissi et al. (2024) combined gamification elements with recommendation features reported increased motivation. However, it lacked a distinct control condition to identify the independent effect of recommendation. For SRL, Du et al. (2024) recruited 81 graduate students from an online course and employed a three-cycle design-based research approach to develop a recommender system for promoting SRL. Through iterative refinement based on interview feedback and pre-post behavioral data, they found that the system significantly improved SRL skills but had no significant impact on academic performance. However, although the ERS developed in this study demonstrated a positive impact on learners’ self-regulation skills, its pedagogical approach notably differs from other EPR systems: it focused on recommending self-regulatory behaviors (such as goal setting or reflection strategies) rather than specific learning materials. Therefore, the findings cannot directly address questions concerning the effects of traditional content-based recommendation strategies on learning processes and outcomes. The question of how high school students respond when recommended with personalized educational content remains insufficiently addressed. In the era when GAI significantly enhances the capabilities of EPR, it is especially important to examine its comprehensive effects on learners’ academic achievement and autonomy, given their rapid growth and increasing practical influence.

\subsection{Educational on-demand help}

On-demand Help (ODH) refers to assistance provided in response to a learner’s proactive request for support, commonly observed in both human teaching interactions and intelligent educational systems (ref). From a Vygotskian perspective, ODH occurs when learners encounter cognitive or practical problems they cannot independently solve, prompting them to seek support from external collaborators (ref). Although seeking help and providing ODH have been longstanding educational phenomena, traditional classroom practices and educational psychology research have mostly focused on teaching methods and classroom activities, paying little attention to how students seek and receive help on an individual level. (Aleven et al., 2003). Some theories have treated "help seeking" as a distinct psychological phenomenon within education, including it as a significant variable within SRL and learner autonomy frameworks (Zimmerman, 2011). Intelligent tutoring systems and digital learning platforms, offer highly controlled environments that enable detailed observation and collection of learners' behavioral and psychological data. This capability facilitates in-depth exploration of help-related behaviors and the underlying learning mechanisms (Yang and Stefaniak, 2023). 

Intelligent tutoring systems supporting ODH enable students to request assistance on challenging content or tasks. Empirical evidence suggests that such systems may improve academic performance. Murphy et al. (2014) conducted a two-year study across nine sites and 20 schools, involving over 70 teachers and approximately 2,000 students each year, to evaluate the implementation of Khan Academy as a supplemental educational resource in classrooms. The study found that while Khan Academy provided students with access to instructional videos and practice exercises, the impact on student outcomes varied significantly acro ss different contexts, and the study did not establish a causal relationship between Khan Academy usage and improved learning outcomes. Similarly, Anderson et al. (1995) developed Cognitive Tutors based on the ACT theory, which provided students with step-by-step hints during problem-solving tasks. Their research indicated that such systems could enhance learning efficiency, with students achieving comparable proficiency levels in less time compared to traditional instruction. Renkl (2002) conducted an experimental study with 60 participants to compare the effects of instructional explanations versus prompting self-explanations, finding that the latter led to better learning outcomes. This suggests that providing instructional explanations during task completion is not always beneficial, especially when compared to encouraging learners to generate self-explanations. Furthermore, Schworm and Renkl (2006) found that in computer-supported example-based learning, providing instructional explanations could reduce students' self-explanation activities, potentially impairing learning. This highlights the need for careful consideration in the design of help systems to balance guidance with promoting active cognitive engagement. These preliminary results indicate that the effects on student development are nuanced and potentially conflicting, highlighting the need for deeper inquiry into the mechanisms at play and the heterogeneous factors that may moderate these effects.

The advent of GAI products, primarily chat-bot-based, significantly enhance the availability and accessibility of ODH. In the context of widespread adoption of GAI for on-demand help in education, research into the educational impacts and underlying mechanisms of GAI-based intelligent learning systems has become critically important (Yan et al., 2024). On the one hand, the default user interaction method of large language model—asking questions to GAIs and getting feedback—fits within the ODH framework. On the other hand, the open, inclusive, and generative characteristics of GAI-powered educational technologies can trigger novel methods, processes and models. Thus, findings previously identified in ODH research are likely to emerge and evolve within GAI contexts. 

Current academic research findings regarding the impact of GAI assistance on education are complex. In professional contexts, Gruda (2024) demonstrated ChatGPT's effectiveness in professional writing assistance. However, contradictions emerge in educational contexts emphasizing human capability growth. For example, Ward et al. (2024) found that the use of AI tools improved academic performance among students by enhancing study habits, time management, and feedback mechanisms. Their study reported a significant reduction in study hours alongside an increase in GPA, suggesting positive academic outcomes. In contrast, research by Bastani et al. (2024) indicated that although GAI instantly enhanced students’ problem-solving performance, these benefits diminished or became negative by final examinations. The experiment involved around 1,000 high school students. The impact of two GPT-4-based AI tutors—one standard (GPT Base) and one with learning safeguards (GPT Tutor)—were tested through the full semester. While both AI tutors enhanced immediate performance during practice sessions, students who used the standard GPT Base performed worse on subsequent assessments without AI assistance, indicating that reliance on GAI can hinder long-term learning, whereas the GPT Tutor's safeguards mitigated these negative effects. The contradictions between short-term versus long-term outcomes and between exam scores versus holistic development could potentially be explained through studies grounded in non-academic variables. For instance, Ward et al. (2024) reported that integrating AI tools positively influenced students' study habits and time management by providing personalized learning support and real-time feedback, leading to improved academic outcomes. Conversely, Zhai et al. (2024) noted a decrease in critical cognitive skills, such as decision-making and analytical reasoning, upon AI integration, attributing this decline to students' over-reliance on AI dialogue systems, which may diminish active cognitive engagement. Furthermore, Darvishi et al. (2024) conducted a randomized controlled experiment with 1,625 undergraduate students across 10 courses to investigate how AI assistance and self-regulation strategies influence the quality of peer feedback. Results indicated that students tended to rely on AI support, and while self-monitoring strategies partially mitigated declines in feedback quality after AI removal, combining self-monitoring with AI prompts did not significantly outperform using AI alone. However, the reported decline in student agency was not directly measured but rather inferred from changes in peer feedback quality indicators. Although important initial findings have emerged, empirical research in this area is still developing, especially in how GAI influences achievement and autonomy in STEM fields across diverse learner populations.

\subsection{Learner autonomy and self-regulated learning}

Learner autonomy (LA) refers to learners' ability to take charge of and manage their own learning processes (Holec, 1981; Little, 1991), involving full responsibility for setting goals, selecting materials, deciding activities and strategies, monitoring progress, and evaluating outcomes. Autonomous learners proactively determine their learning content and methods, take complete responsibility for the learning processes and outcomes, seek assistance without complete dependency, and adjust strategies according to learning needs and contexts. Recent theoretical developments divide LA into technological, psychological, political, and social dimensions (Bensen, 2001). Among these, the technological and psychological dimensions align closely with the concept of SRL, as both describe learner control and management over the learning process, including motivation, goal setting, execution monitoring, and reflective evaluation. The political and social dimensions of LA emphasize learners' autonomous beliefs in relationships and responsibilities, incorporating learning environments, teacher-student relationships, and broader political contexts within education and society. Specifically, the political dimension addresses learners' perceptions of freedom, rights in educational decision-making, and their relationships with educational authorities. The social dimension examines students’ tendency to form interdependent learning relationships with those around them, including their inclination to communicate with, seek support from, and compare themselves to their teachers and classmates (Murase, 2015).

Autonomous learners actively manage their learning through self-regulated strategies. SRL is a crucial process that enables learners to actively set goals, regulate their cognitive, motivational, and behavioral strategies, and adapt to different learning environments (Zimmerman, 2011). Models including Zimmerman’s cyclical model, Pintrich’s self-regulation framework, and Winne and Hadwin’s information processing model, emphasize the dynamic nature of SRL, highlighting its role in goal setting, self-monitoring, and feedback-based adjustments (Zimmerman, 2011; Pintrich, 2000; Winne \& Hadwin, 1998). Despite minor variations in perspectives, these models share a common focus on the interplay between cognition, motivation, and behavioral control as key components of SRL. Self-regulation is regarded as a crucial ability fostering both short-term academic success and long-term developmental achievements (McClelland \& Cameron, 2012). In the field of STEM learning, students who develop SRL skills perform better on immediate tasks, such as problem-solving and computational thinking, and retain these skills over time, improving their ability to adapt and succeed in complex academic tasks (Tsai et al., 2013). Self-regulation interventions have also been shown to exert lasting influence beyond academic settings, with significant effects on individuals’ lifelong health and social outcomes (Pandey et al., 2018).

In terms of measurements, numerous tools have been developed for SRL, including questionnaires, interviews, think-aloud protocols, and online-based assessment systems (Wen et al., 2023). However, due to its origins in social cognitive theories, the idea of SRL primarily focuses on the behavioral side of a learners’ autonomous tendencies. SRL theories acknowledge the existence of factors regarding individual autonomous beliefs and social-political perspectives but typically treats these factors as backgrounds. Given that education aims not only at developing specific behavioral capabilities but also at cultivating freewill and autonomous personalities, measuring self-regulatory behaviors alone does not necessarily translate into the broader concept of autonomy. Utilizing questionnaires based on LA theory to examine autonomy’s all four dimensions in technological, psychological, political, and social aspects can offer a more comprehensive view of the impact of AI on learner autonomy.

Recent empirical studies have illuminated the relationship between students' academic achievement levels and their SRL strategies and autonomy, suggesting differentiated instructional approaches may be beneficial. Cheng et al. (2024) conducted a study involving 66 secondary school students and 59 university students, utilizing trace data to analyze SRL processes during multi-source writing tasks. Findings indicated that lower-performing secondary students predominantly engaged in basic SRL processes like orientation, relying heavily on task instructions and rubrics, whereas higher-performing students employed more advanced strategies such as re-reading and organization, highlighting a need for scaffolding to enhance SRL skills among lower achievers. In higher education, Oguguo et al. (2023) surveyed 256 undergraduates to assess the impact of SRL, autonomy, and agency on academic achievement in an online research methods course. The study revealed a strong positive correlation (r = 0.862) between these variables and academic performance, with SRL, autonomy, and agency collectively accounting for 74.4\% of the variance in students' achievement, underscoring the importance of fostering these skills for academic success. Regarding the use of GAI, in a randomized controlled trial, Li et al. (2025) developed and tested GAI-powered SRL hints—adaptive prompts generated by GAI based on real-time analytics of learners' SRL processes. Similar to Du et al. (2024), the design stressed on the metacognition aspect of learning instead of the cognitive (i.e. knowledge comprehension) one. The study found that participants receiving these personalized SRL hints exhibited significantly more metacognitive learning behaviors compared to control groups. Collectively, these studies suggests that lower-achieving students might benefit from more structured learning support to develop SRL and autonomy, and the latter skills can be fostered through hints and guidance generated by GAI. However, existing intervention studies on GAI-powered feedback tend to focus exclusively on SRL or autonomy, often without reporting corresponding changes in achievement. Conversely, studies that report achievement gains typically do not directly measure SRL or autonomy, instead inferring their impact through indirect reasoning. This disconnect highlights the need for systematic investigations into the interplay between GAI usage patterns, achievement and autonomy outcomes.

\subsection{Research questions}

Building upon the reviewed literature, it is evident that GAI hold great potential for enhancing educational outcomes through personalized support and adaptive interventions. However, there remains limited understanding of how usage patterns of GAI‑powered feedback impact students across various academic achievement levels. To address this gap, the present study proposes three research questions.

RQ1: How does compulsory usage of GAI-powered personalized feedback affect achievement and autonomy across different academic achievement groups? 

RQ2: How does autonomous usage of GAI-powered personalized feedback impact achievement and autonomy across various academic achievement groups? 

RQ3: Integrating findings from RQ1 and RQ2, are there optimal usage patterns tailored to students from different academic achievement groups?

\section{Methodology}

\subsection{Experiment 1: Compulsory usage of GAI-powered personalized recommendation}

Experiment 1 explored whether providing personalized recommendations with AI-powered hints influences students' achievement and autonomy within an authentic school setting (RQ1). High-school physics was chosen as the intervention discipline in control for subject-related variability. Participants were individually randomized into three experimental conditions: personalized recommendations with GAI-powered hints (Group A), personalized recommendations with answers and explanations from workbooks (Group B), and the control group (Group C1).

\subsubsection{Participants and sampling}

The participants in this study were students recruited from a high school in southwest China. Prior to implementation, the experimental protocol was reviewed and approved by the school’s academic and administrative teams and received ethical approval from the Institutional Review Board of Tsinghua University. Participants' rights to informed consent and voluntary withdrawal from the study were fully guaranteed throughout the research process.

In total, 160 participants were recruited from three Grade 10 classes, representing a broad range of absolute academic achievement groups (pre-test scores ranging from 16 to 90 out of 100), ensuring substantial heterogeneity in the sample. A stratified random sampling strategy was adopted. Within each of the three classes, individual participants were randomly assigned to three conditions: AI-powered content recommendations (Group A, n = 54), conventional instructional content recommendations (Group B, n = 53), and no recommendations (Group C1, n = 53). The number of participants allocated to each condition was evenly distributed within each class. Given the commonly large intra-class correlation, such stratified sampling strategy ensures variation under each condition. 

\subsubsection{Procedures}

All participants simultaneously received a five-week experimental intervention after providing informed consent. Throughout the intervention period, all participants (including those in the control group) were required to consistently mark and submit their errors in daily physics homework on answer sheets. This approach allowed the recommendation system to collect students' learning data to generate personalized recommendations, while keeping participants blank of their group assignment to avoid John Henry effect and Hawthorne effect. This process is conducted based on the routine school practices, which require students to self-check their homework, report the errors, and self-correct their homework.

After daily error report, participants in treatment group A and B received individualized review materials recommended by the system over the subsequent days. Specifically, participants in Group A received recommended exercise problems, GAI-powered solution hints for these problems, and corresponding answers and explanations from workbooks. In contrast, participants in Group B received only recommended exercise problems along with their answers and explanations from workbooks, without GAI-powered solution hints. Both groups were encouraged to voluntarily utilize these recommended resources during their independent study time. Group C is the control group where participants received no recommendations.

\begin{figure}
    \centering
    \includegraphics[width=0.7\linewidth]{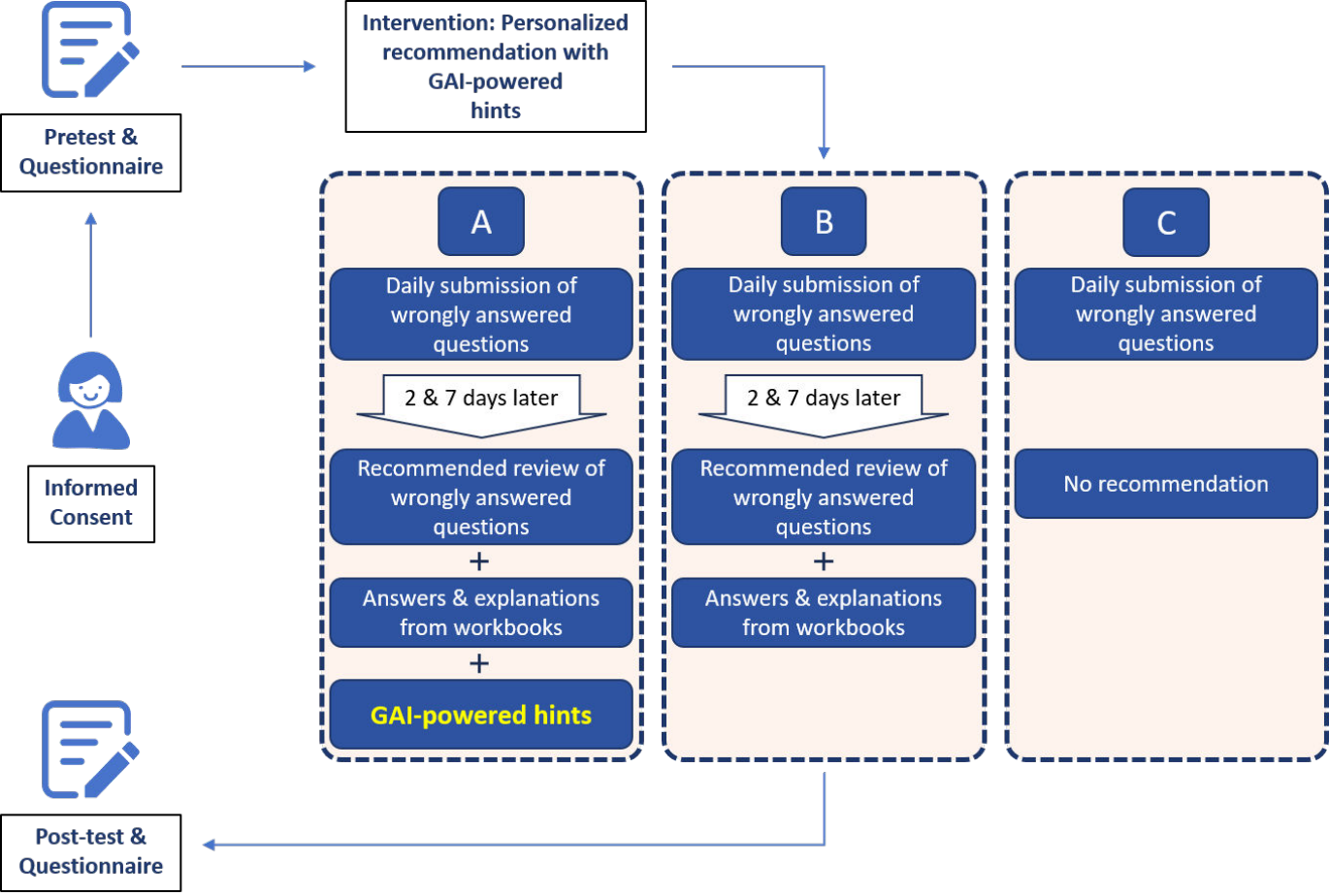}
    \caption{Experiment Procedure}
    \label{fig:procedure-1}
\end{figure}

Before and after the intervention, researchers administered SRL questionnaires. Pre- and post-intervention test scores were used as outcome indicators of academic achievement in subsequent analyses. The procedure is displayed in Figure \ref{fig:procedure-1}.

\subsubsection{Instrument design}

\paragraph{GAI-powered hints}

\begin{figure}
    \centering
    \includegraphics[width=0.9\linewidth]{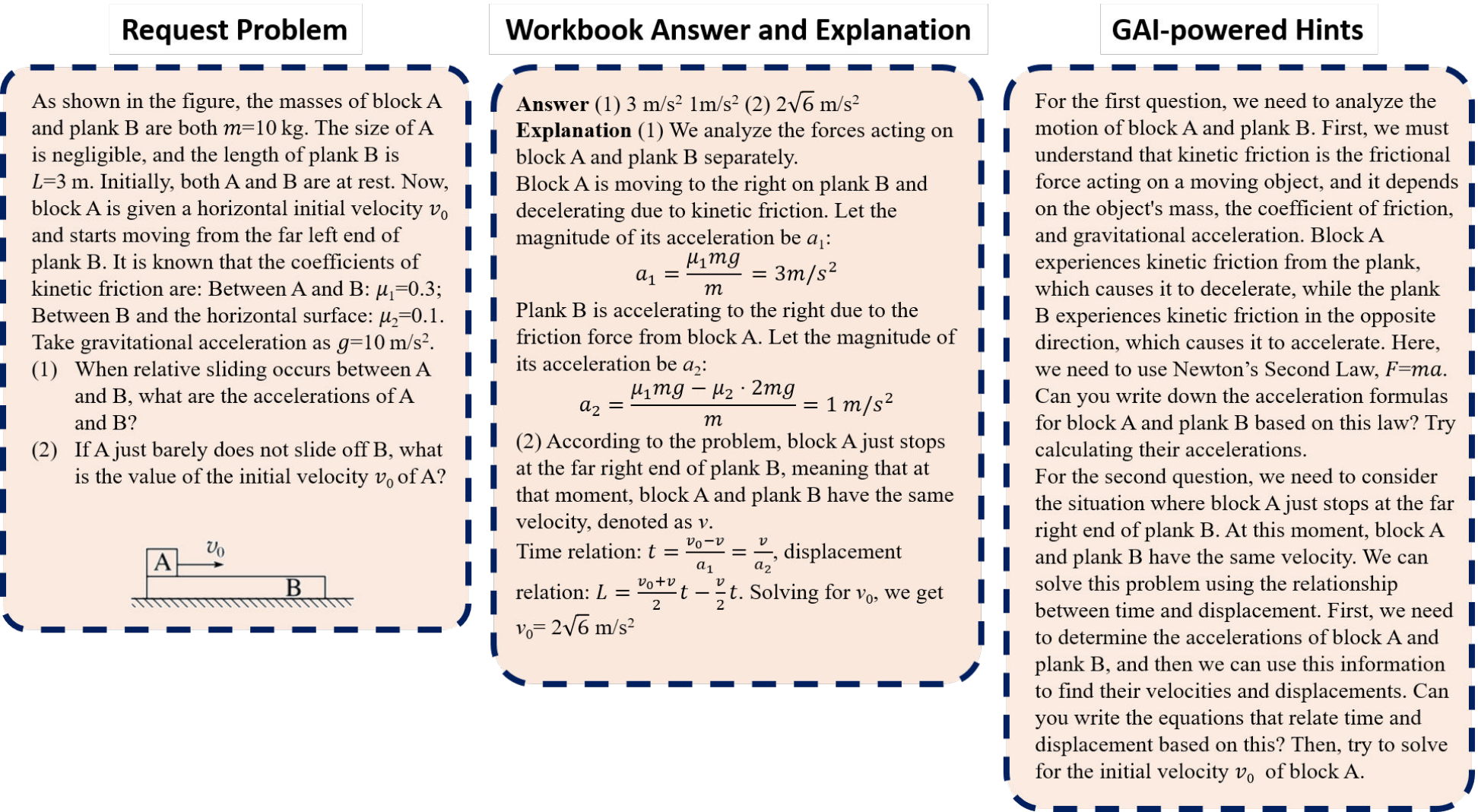}
    \caption{An example of GAI-powered hints and answers and explanations from workbooks}
    \label{fig:hint-eg}
\end{figure}

This study developed a GAI workflow leveraging comprehensive retrieval augmentation and multi-agent techniques to generate heuristic solution hints for recommended problems. These hints were provided alongside recommended practice problems exclusively to participants in Group A, serving as scaffolding support when students encountered difficulty during problem-solving. Compared to conventional workbook instructional answers and explanations, these AI-generated hints were characterized by personalized diagnosis, emotional support, and heuristic rather than direct answer provision. A typical example of the GAI-powered hints and the corresponding answer and explanation from workbook is displayed in Figure \ref{fig:hint-eg}.

Regarding personalized diagnosis, the workflow identified students' specific mastery levels in declarative knowledge comprehension and procedural knowledge application based on their incorrect answers or solution procedures, combined with the correct solution process and relevant knowledge points. For personalized explanation, the workflow tailored the generated hints according to the diagnostic outcomes, concisely reviewing areas where students demonstrated sufficient understanding while elaborating or emphasizing content in areas where they showed weaknesses (Bloom, 1971).

In terms of emotional support, the study observed that typical workbook answers and explanations were mostly neutral and objective, whereas GAI could offer feedback with greater emotional resonance through carefully crafted instructions and example-based prompting. Providing emotionally supportive feedback has been shown to facilitate learners' cognitive and emotional development, including using non-judgmental language when pointing out mistakes, affirming students' existing efforts and understanding, encouraging further attempts, and promoting an accepting attitude toward errors (Rogers et al., 2011). These principles were explicitly integrated into the workflow’s prompts, and chain-of-thought examples (Wei et al., 2022) were provided for agent training and reference.

Regarding heuristic feedback, the workflow intentionally refrained from directly providing numerical solutions, multiple-choice answer analyses, or explicit computational formulas. Instead, it offered fundamental conceptual principles, contextual analysis, and directional suggestions for problem-solving attempts, thereby providing optimal scaffolding for independent learning (Kirschner et al., 2006). 

To evaluate diagnostic accuracy, the study conducted comparative assessments of the workflow against human teachers. The quality of generated hints was also monitored and validated through participant ratings. 

The reliability of GAI hint-generating workflow was evaluated through an offline experiment. Ten high school physics teachers, all graduates of China’s top universities, were invited to provide feedback on ten flawed student responses. The same task was completed independently by the GAI workflow across ten iterations. In a subsequent blind review, three expert teachers were asked to evaluate the quality of the feedback without knowing its source (human or AI). Results from independent samples t-tests revealed no significant differences between the human-generated and AI-generated feedback.

During the intervention, students' subjective experiences were also collected and analyzed. Participants who received GAI-powered hints were invited to rate the quality of the AI-generated feedback. Three response options were provided: "very helpful," "somewhat helpful," and "not helpful." A total of 112 voluntary ratings were collected. Among them, 82 ratings (73.2\%) indicated "very helpful," 28 ratings (25\%) indicated "somewhat helpful," and 2 ratings (1.8\%) indicated "not helpful." These results suggest that the GAI-powered hints used for intervention in this study achieved a high level of user satisfaction.

\paragraph{Recommendation rule}

This study designed an easy-to-implement recommendation rule grounded in the principle of spaced repetition (Ebbinghaus, 1913; Dehaene, 2020) and prevalent practices in secondary education: after learners submit their errors in daily homework, they receive repeated recommendations of previously incorrectly answered problems on the second- and seventh-days following submission.

Review of incorrectly answered problems are widely advocated in Chinese high schools, since these items often signify a student's "zone of proximal development"—tasks that learners cannot yet perform independently but can accomplish with appropriate assistance (Vygotsky, 1978). Practicing within this zone is generally regarded as highly beneficial for cognitive growth. However, despite the recognized value of reviewing wrongly answered problems, practical implementation remains technically ambiguous and challenging (e.g., complicated operation, difficulty in monitoring and assessment), hindering educators' effective adoption of this practice. Integrating an error-report and recommendation system into routine classroom activities could significantly alleviate these implementation barriers. The approach adopted in the current study exemplifies such integration.

Drawing from the principles of spaced repetition, this study’s recommendation system provided learners with feedback regarding their submitted homework on the second and seventh days following the original submission. In terms of knowledge retention and skill training, feedback delivered at increasingly spaced intervals has been shown to outperform immediate feedback (Cepeda et al., 2006). Guided by principles of retrieval practice, the recommended problems, AI-generated solution hints (Group A only), and conventional instructional answers and explanations (both Group A and B) were presented separately on distinct feedback sheets with clear instructional prompts. Specifically, for Group A participants, the instructional prompts encouraged them first to attempt solving the problems independently, then refer to the AI-generated solution hints only if they encountered difficulties and finally consult the provided answers and explanations if still unable to resolve the problem. For Group B participants, prompts encouraged independent problem-solving attempts first, followed directly by reference to conventional instructional answers and explanations if needed.

\paragraph{Workbook answers and explanations}

All classes within the participating schools’ entire grade level utilized the same set of supplementary instructional materials for homework assignments. Accordingly, researchers obtained the corresponding answers and explanations provided by these materials. When recommending specific problems, the system simultaneously supplied these publisher-authored answers and explanations to students in both Group A and Group B. These conventional materials, compiled by publishers, typically presented solutions by first stating the final answer, followed by a brief explanation. Although these explanations were limited in depth and comprehensiveness, they nonetheless represented the primary form of learning scaffolding for most Chinese high school students.

To maximize the effectiveness of system-recommended exercises and the utilization of GAI-powered hints, workbook answers and explanations were provided separately from other materials during the intervention. Specifically, students in Group A received three distinct booklets for each assignment: (a) a booklet containing only the exercise questions, (b) a booklet with GAI-powered hints corresponding to those questions, and (c) a booklet with the official answers and detailed explanations. Prior to the intervention, Group A participants were provided with a written instruction manual and received oral reinforcement of the guidelines. They were explicitly instructed to first attempt the problems independently using the first booklet. For items they found challenging or were uncertain about, they were directed to consult the second booklet for GAI-powered hints, integrate the information, and then attempt the questions again. Only after these steps, for questions they still could not solve or for which they remained uncertain, were they permitted to use the third booklet to check the answer and review the explanation.

In contrast, students in Group B received only the first and third booklets and did not have access to GAI-powered hints. They were provided with the same instructional philosophy: to attempt the exercises independently first, and, if difficulties arose, to consult the answer and explanation booklet for guidance and verification. This design ensured that both groups followed a structured problem-solving process, with the critical difference being the availability of GAI-powered hints in Group A. This arrangement allowed for an examination of the impact of such hints on students’ academic achievement and learning autonomy.

\paragraph{Paper-based interaction}

To comply with the school's management policy prohibiting the use of smart electronic devices in classrooms, and to ensure the intervention design's scalability and transferability across different schools, this study implemented all interventions using paper-based interactions. The system employed scanning and optical character recognition technologies to capture and analyze the submitted problem-solving attempts. All recommended and generated contents were printed by the school's printing office and subsequently distributed to students.

\subsubsection{Measurements}

\paragraph{Autonomy}

Experiment 1 used a 15-item SRL questionnaire developed by Dowson and McInerney (2004) as instrument for autonomy. The reliability of the instrument was confirmed, with Cronbach's alpha coefficients of 0.790 for the pretest and 0.823 for the post-test. 

\paragraph{Academic achievement}

The experiment utilized students' exam scores as pre- and post-intervention measures of academic achievement. Pre-intervention data included students' final exam scores from the semester preceding the experiment. The post-intervention measure was the final exam score obtained after completing the intervention period. Before conducting regression analyses, all test scores were converted into standardized scores (z-scores).

\paragraph{User experience}

During the experimental period, researchers conducted interviews with 30 participants based on maximum variation sampling to gain insights into their personal experiences with the system and its integration with other learning tasks. The interviews were semi-structured to capture a broad range of potential insights. Key interview questions included:

(1) How would you evaluate the quality of feedback provided by the system?

(2) How did you utilize the recommended materials provided by the system?

\subsubsection{Data analysis}

The reliability analysis for questionnaire data was performed using SPSS 26.0, and regression analyses were conducted using STATA 17. To address the first research question, the two experimental conditions (Group A and Group B) were included in the same regression model as two distinct binary variables, each compared separately against the control condition. An F-test was then applied to evaluate whether the effect sizes significantly differed between the two groups.

The overall treatment effect was estimated through regression analysis across the entire sample. To explore heterogeneity, the full sample was divided into three subgroups based on their pretest academic achievement (top third, medium third, bottom third). Separate linear regressions were performed for each subgroup. In addition to the experimental grouping variables, pretest academic performance, pretest self-regulation level, and class fixed effects were included as control variables in the regression models to enhance overall model fit.

The regression equation was specified as follows:

\begin{equation}
\begin{aligned}
Outcome_{ij}=&\beta_{10}+\beta_{11}TreatmentA_{ij}+\beta_{12}TreatmentB_{ij}+\beta_{13}Pretestscore_{ij}+ \\
&\beta_{14}PreSRL_{ij}+\sum_j{\gamma_{1j}Class_j}+\epsilon_{ij}
\end{aligned}
\end{equation}

where $Outcome_{ij}$ is the standardized exam grade or SRL questionnaire result of student $i$ in class $j$; $TreatmentA_{ij}$ and $TreatmentB_{ij}$ are binary variables indicating treatment assignments for student $i$ in class $j$; $Pretestscore_{ij}$ controls for student performance on physics, capturing the exam score in the previous term of student $i$ in class $j$; $PreSRL_{ij}$ controls for the SRL self-report questionnaire score gathered before the experiment of student $i$ in class $j$; and Class $j$ controls class $j$’s class level fixed effect. Robust standard errors are calculated at the individual level, which is the unit of randomization.

\subsection{Experiment 2: autonomous usage of GAI-powered on-demand help}

In the semester following Experiment 1, Experiment 2 was conducted within the same grade at the same school to address research questions 2 and 3. Experiment 2 included two experimental groups—on-demand help with full student control and on-demand help with student-system shared control--and a non-intervention control group to examine how fully learner-controlled on-demand help and algorithm-learner jointly controlled help-seeking, under AI-generated content scenarios, influence students' academic performance and SRL.

Fully learner-controlled on-demand help allowed students to identify specific topics or problems for which they needed assistance (e.g., "I don't understand Newton's laws" or "I got this question wrong") and to specify the type and content of help they desired. Students could freely select services such as heuristic solution hints, explanations of concepts, or additional relevant practice problems. Conversely, the algorithm-learner jointly controlled help-seeking permitted students only to identify problematic topics or questions, after which the system autonomously recommended appropriate content based on its algorithm. For instance, if a student raised questions about only one problem within the "Newton's second law" chapter, the system would provide heuristic solution hints only for that problem. However, if multiple related issues were identified, the system inferred substantial gaps in foundational knowledge and thus provided both explanations of concepts and additional relevant practice problems.

\subsubsection{Participants and sampling}

In Experiment 2, participant recruitment expanded to include all students in the grade level enrolled in physics courses (n = 373), following the same informed consent and ethical principles as in Experiment 1. Among the recruited students, 124 were assigned to the on-demand help group where students have full control over feedback type selection (Group D), 124 to the shared-control help group where student can raise questions but the system decides which types of feedback are provided (Group E), and 125 to the control group (Group C2). Participants were individually randomized within each of the seven administrative classes, ensuring equal distribution among the three experimental conditions within each class. Ultimately, 266 students (Group D: n = 87; Group E: n = 88; Group C2: n = 91) provided valid pre- and post-intervention test scores and completed the pre- and post-SRL questionnaires.

\subsubsection{Procedures}

\begin{figure}
    \centering
    \includegraphics[width=0.9\linewidth]{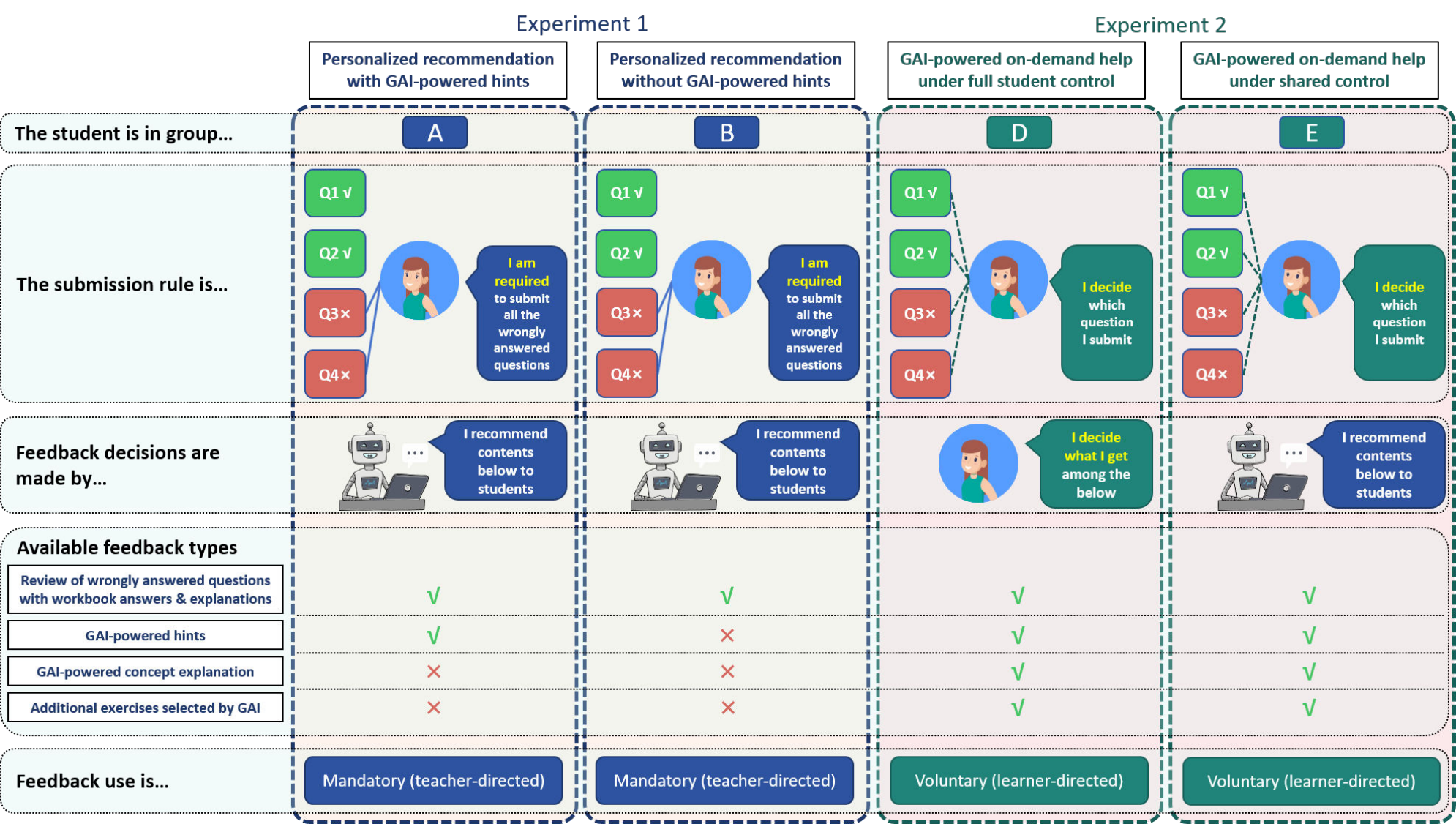}
    \caption{Interventions among groups}
    \label{fig:intervention}
\end{figure}

Upon commencement of the experiment, all participants received a loose-leaf notebook. For students in Groups D and E, the notebook included detailed instructions for using the physics learning support system, along with loose-leaf sheets like the submission sheets employed in Experiment 1. Detailed instruments are shown in Figure \ref{fig:intervention} and the following section. Participants marked problems they answered incorrectly or found difficult on these sheets, scanned them, and uploaded them to the researcher-developed system, subsequently receiving feedback the next day. Students in Group C2 received identical notebooks, though without physics-related materials.

This experiment lasted for five weeks. Researchers administered identical questionnaires and collected academic performance data at two points: immediately before and after the intervention period.

\subsubsection{Instrument design}

\paragraph{GAI-powered on-demand help system with full student control of feedback}

This system was accessible exclusively to students in Group D. Participants could specify problems they struggled with and outline explicit requests. Without such detailed requests, the students would not receive effective feedback. Requests could include solution hints, explanations of concepts, or additional relevant practice problems. Students submitted their worksheets during daily evening self-study sessions; teachers then scanned and uploaded these sheets to the server for system analysis. On the following day, each student who provided clear requests received personalized AI-generated feedback addressing their specific needs.

Solution hints were generated using the same workflow and principles as in Experiment 1. For concept explanations, an AI-agent-based workflow used large language models to provide clear, concise, and vivid explanations of relevant physics concepts, laws, and formulas based on student-reported questions or described knowledge needs. These explanations were contextualized with real-life examples to facilitate deeper understanding (Barron \& Darling-Hammond, 2008). When specific questions were provided, the workflow identified and explained each of the key concepts contained in the question individually, subsequently highlighting the connections between these concepts and the problem-solving process without explicitly providing detailed solutions or final answers.

Regarding the additional relevant practice problems, students could either submit a sample problem or describe their practice needs. The AI workflow retrieved related practice problems from a carefully curated question bank containing approximately 6,000 high-quality physics problems, each including text, illustrations, answers, detailed explanations, and metadata such as difficulty level and associated knowledge points. The system utilized OpenAI's embedding models to transform the question bank into a vector database. Upon identifying student requests, the workflow input both the original sample problem (or described need) and the associated AI-generated labels into the vector retrieval engine to produce multiple potential matches. Subsequently, a dedicated selection and deduplication agent evaluated these matches, selecting and delivering the one or two most appropriate practice problems to the student.

\paragraph{GAI-powered on-demand help system with shared control of feedback}

This system was made accessible exclusively to students in Group E. Participants could indicate specific problems they encountered difficulties with but were not required to specify explicit needs. Based on a recommendation algorithm, the system autonomously provided students with feedback tailored to their identified problems. The range of available feedback was identical to that provided to Group D, including solution hints, concept explanations, and additional relevant practice problems.

The system proposes a feedback recommendation algorithm grounded in students’ temporal submission behaviors. Like Huptych et al.’s approach (2017), the algorithm introduces a learning factor ($LF$) as a quantitative indicator of a student’s current learning condition, incorporating both the frequency and recency of their submissions. The system counts the number of submissions at different time and applies an exponential decay function to assign greater weights to more recent activities, ensuring that recent behaviors have a stronger influence on $LF$, which is computed as follows:

\begin{equation}
\begin{aligned}
    LF=3\times\left(\frac{2}{3}\right)^D+\sum_{i=1}^n{c_i\times\left(\frac{2}{3}\right)^{d_i}}
\end{aligned}
\end{equation}

where $D$ denotes the number of distinct days on which submissions occurred, $c_i$ is the number of submissions on day $i$, and $d_i$ is the number of days elapsed since that submission. Based on the computed $LF$, the system adopts a tiered feedback strategy to support individualized learning. When the factor is small ($LF < 5$), students receive both additional relevant practice questions and their solution hints are provided for each help request, aiming to deepen conceptual understanding and support transfer. For moderate values ($5 \leq LF < 10$), only additional exercises are provided, supporting students consolidate skills through continued practice. In contrast, a high $LF$ value ($LF \ge 10$) suggests insufficient mastery. In such cases, the system randomly selects either an additional relevant practice problem or a piece of concept explanation as feedback for each help request, in addressing potential learning gaps.

\paragraph{Loose-leaf notebook-based interaction}

The paper-based interaction approach employed in Experiment 1 was continued and refined in Experiment 2. Based on user interviews conducted after Experiment 1, which highlighted issues such as the inconvenience of organizing scattered feedback sheets, both submission sheets and feedback sheets were transitioned from standard A4 paper to loose-leaf A4 paper. Additionally, the researchers provided loose-leaf notebooks to participants across all experimental conditions, enabling students to conveniently organize and manage their feedback sheets. 

\subsubsection{Measurements}

\paragraph{Autonomy}

In the interview analysis of Experiment 1, the researchers found that, beyond self-regulated behaviors themselves, students’ perceptions of teacher authority and teacher–student relationships also substantially influenced their learning behaviors. Therefore, in Experiment 2, the study incorporated a more comprehensive learner autonomy questionnaire to further examine whether GAI affects students’ development not only at the behavioral level but also at the political and social dimensions.

Experiment 2 employed a Chinese-translated LA questionnaire, originally developed by Murase (2015) for English learners, and adapted it to fit for STEM context. The adaptation involved no item deletions, only modifications from English-specific expressions to general academic contexts, along with translation. A pilot study among 500 Chinese university students indicated high reliability across the four dimensions (technological, psychological, political, and social), with Cronbach's alpha coefficients of 0.991, 0.997, 0.999, and 0.998, respectively. Reliability analysis of the questionnaire with Experiment 2 participants yielded Cronbach's alpha values of 0.901, 0.866, 0.720, and 0.831 across these dimensions, respectively. 

\paragraph{Academic performance}

Experiment 2 also used exam scores from the school's regular academic schedule as pre- and post-intervention measures. The pretest data comprised students' final examination scores from the semester immediately preceding Experiment 2. Post-test data consisted of final exam scores at the conclusion of the intervention period. Prior to regression analysis, all performance scores were converted to standardized z-scores.

\paragraph{User experience}

In the later stages of the experiment, researchers conducted rolling interviews with 15 participants selected according to the principle of maximum variation sampling. The purpose was to understand participants' system usage experiences and perceptions, as well as to explore potential underlying mechanisms driving experimental results. Interview questions included:

(1) Do you think the feedback provided by the system was helpful to you?

(2) How did you make use of the materials provided by the system?

\subsubsection{Data analysis}

Experiment 2 employed the same analytical software (SPSS and STATA 17) and modeling strategies as Experiment 1. To address RQ2 and RQ3, the two experimental conditions (Groups D and E) were included as two distinct binary variables in the same regression model, each compared separately with the control group. F-tests were subsequently performed to evaluate whether significant differences in effect sizes existed between the groups. Regression analyses were conducted on the whole sample as well as the top, medium, and bottom third of the sample by pretest score.
The regression model was specified as follows:

\begin{equation}    
\begin{aligned}
Outcome_{ij}=&\beta_{20}+\beta_{21}TreatmentD_{ij}+\beta_{22}TreatmentE_{ij}+\beta_{23}Pretestscore_{ij}+ \\
&\beta_{24}PreLA_{ij}+\sum_j{\gamma_{2j}Class_j}+\epsilon_{ij}
\end{aligned}
\end{equation}

where $Outcome_{ij}$ is the standardized exam grade or SRL questionnaire result of student $i$ in class $j$; $TreatmentD_{ij}$ and $TreatmentE_{ij}$ are binary variables indicating treatment assignments for student $i$ in class $j$; $Pretestscore_{ij}$ controls for student performance on physics, capturing exam score in the previous term of student $i$ in class $j$; $PreLA_{ij}$ controls for the SRL self-report questionnaire score gathered before the experiment of student $i$ in class $j$; and $Class_j$ controls class $j$’s class level fixed effect. Robust standard errors are calculated at the individual level, which is the unit of randomization.

\section{Results}

\subsection{Experiment 1: compulsory usage of GAI-powered personalized recommendation}

\subsubsection{Sample}

Among all participants, 121 (Group A: n = 42; Group B: n = 38; Group C1: n = 41) submitted valid pre- and post-test scores and completed the SRL questionnaires. Students were grouped into top-, medium-, and bottom-third cohorts according to their pre-test performance. Group A contained 15 top students, Group B contained 12 top students, and Group C1 contained 14 top students. Fourteen medium-third students were randomly distributed to Group A, 10 to Group B, and 17 to Group C1. Thirteen bottom-third students were distributed to Group A, 16 to Group B, and 10 to Group C1.

Across the entire sample, one-way ANOVA results revealed no statistically significant differences among Groups A, B, and C1 in terms of pretest academic performance, F(2, 118) = 0.09, p = .914, or pretest SRL, F(2, 118) = 1.93, p = .150. Consistent patterns were observed within each performance-level subsample. Specifically, no significant group differences were found in pretest academic performance within the top-third subsample, F(2, 38) = 1.67, p = .202; the medium-third subsample, F(2, 38) = 0.81, p = .453; or the bottom-third subsample, F(2, 36) = 0.74, p = .484. Likewise, no significant differences were observed in pretest SRL within the top-third subsample, F(2, 38) = 1.90, p = .163; the medium-third subsample, F(2, 38) = 0.08, p = .928; or the bottom-third subsample, F(2, 36) = 0.71, p = .498. These findings suggest that the three groups were comparable at baseline across both academic and self-regulatory domains.

The assumption of homogeneity of variance for both pretest academic performance and pretest SRL was met across all groups and performance-level sub-samples, as indicated by non-significant Bartlett’s tests (p > .05 for all comparisons). These results, combined with the non-significant ANOVA findings, indicate that the three experimental groups were well-matched at baseline, demonstrating strong comparability and equivalence both in the overall sample and within each sub-sample prior to the intervention.

\subsubsection{Experiment implementation}

Among the 42 participants in Group A, 38 actively engaged with the system, whereas 31 out of 38 participants in Group B used the system. The relatively high system usage rate demonstrates the successful implementation of the personalized recommendation, in which teachers require students to submit incorrectly answered problems to the system and utilize the feedback in daily homework assignments. Among students who interacted with the system, the average number of wrongly answered question submission was 11.4 (SD = 7.9) for Group A and 8.3 (SD = 7.0) for Group B. A one-way ANOVA indicated no significant difference in the average number of system interactions between the two groups (F(1, 68) = 2.94, p = 0.091). Regarding achievement subgroups, students in the top-, medium- and bottom third submitted 7.2 (SD = 5.6), 11.3 (SD = 7.6) and 9.7 (SD = 8.1) wrongly answered questions. For each submitted wrongly answered question, a student would receive one recommendation on the second day after submission, and the same recommendation repeated on the seventh day.

\subsubsection{Results on test scores}

As shown in Table \ref{achive-1}, regression analyses on the whole sample showed no significant impact of either intervention on academic performance. However, subgroup analyses revealed that for low-performing students, the GAI-powered recommendation significantly improved performance by 0.673 SD (p<0.05), and traditional recommendations had a marginally significant effect of 0.743 SD (p=0.051), both reflecting medium-to-large effect sizes. Conversely, neither intervention significantly affected high-performing students. Unexpectedly, recommendations with workbook answer and explanation significantly reduced performance among the medium-third subgroup (d=-0.539, p<0.05), while the AI intervention had no such negative effect (d=0.101, p>0.05). An F-test confirmed this significant difference in effects between the two interventions (F (1,34) =5.68, p<0.05), suggesting giving AI-generated heuristic solution hints is better than solely providing traditional feedback.

\begin{table}[htpb]
  \centering
\begin{threeparttable}
\caption{Treatment Effects on Achievement in Experiment 1}
  \label{achive-1}
\begin{tabular}{lrrrr}
\toprule
              & \multirow{2}{*}{Whole sample} & \multicolumn{3}{c}{Sub-sample by achievement} \\
              &                               & \multicolumn{1}{c}{Top}           & \multicolumn{1}{c}{Medium}        & \multicolumn{1}{c}{Bottom}        \\
              & \multicolumn{1}{c}{(1)}                            & \multicolumn{1}{c}{(2)}             & \multicolumn{1}{c}{(3)}           & \multicolumn{1}{c}{(4)}            \\
              \midrule 
Treatment A   & 0.212                         & 0.107         & 0.101         & 0.673*        \\
              & (0.154)                        & (0.241)        & (0.277)        & (-0.315)        \\
Treatment B   & 0.051                         & -0.043        & -0.539*       & 0.743         \\
              & (0.181)                        & (0.248)        & (0.246)        & (-0.367)        \\
Pretest score & 0.602*                        & 0.733**       & 0.325         & 0.35          \\
              & (0.093)                        & (0.196)        & (0.539)        & (0.414)        \\
Pretest SRL   & -0.073                        & -0.123        & -0.008        & -0.006        \\
              & (0.155)                        & (0.223)        & (0.274)        & (0.483)        \\
              \midrule
N             & 121                           & 41            & 41            & 39            \\
$R^2$            & 0.56                          & 0.48          & 0.29          & 0.25          \\
F             & 25.43***                      & 6.52***       & 3.78**        & 5.84***       \\
p             & 0.000                             & 0.000             & 0.005         & 0.000            \\
\bottomrule
\end{tabular}
\begin{tablenotes}
\item Notes: * p < 0.05, ** p < 0.01, *** p < 0.001; standard errors in parentheses. 
\end{tablenotes}
\end{threeparttable}
\end{table}

\subsubsection{Results on autonomy (self-regulated learning)}

According to Table \ref{srl-1}, using SRL level as the dependent variable, interventions A and B produced marginally significant negative effects of 0.279 (p = 0.091) and 0.379 (p = 0.071) standard deviations, respectively, for the full sample. Analysis of sub-samples of the top, medium, and bottom thirds students revealed that GAI-powered hint recommendation and workbook answer and explanation recommendation significantly reduced the post-test SRL levels of high-scoring students by 0.477 (p < 0.05) and 0.973 (p < 0.05) standard deviations, respectively, whereas no significant effects were observed in the other sub-samples. Furthermore, no significant difference in effect sizes was detected between the two interventions.

\begin{table}[htpb]
  \centering
\begin{threeparttable}
\caption{Treatment Effects on Self-regulated Learning in Experiment 1}
  \label{srl-1}
\begin{tabular}{lrrrr}
\toprule
              & \multirow{2}{*}{Whole sample} & \multicolumn{3}{c}{Sub-sample by achievement} \\
              &                               & \multicolumn{1}{c}{Top}           & \multicolumn{1}{c}{Medium}        & \multicolumn{1}{c}{Bottom}        \\
              & \multicolumn{1}{c}{(1)}                            & \multicolumn{1}{c}{(2)}             & \multicolumn{1}{c}{(3)}           & \multicolumn{1}{c}{(4)}            \\
              \midrule
Treatment A   & -0.279                        & -0.477*        & -0.169       & -0.312        \\
              & (0.164)                       & (0.218)        & (0.324)      & (0.346)       \\
Treatment B   & -0.379                        & -0.973*        & 0.003        & -0.193        \\
              & (0.207)                       & (0.432)        & (0.449)      & (0.315)        \\
Pretest score & -0.024                        & 0.141          & -1.289*      & 0.219         \\
              & (0.097)                       & (0.324)        & (0.533)      & (0.368)        \\
Pretest SRL   & 1.106***                      & 1.189***       & 0.962*       & 1.467***      \\
              & (0.194)                       & (0.252)        & (0.375)      & (0.324)        \\
              \midrule
N             & 121                           & 41             & 41           & 39            \\
$R^2$            & 0.3                           & 0.4            & 0.27         & 0.42          \\
F             & 8.37***                       & 6.24***        & 2.92*        & 5.78***       \\
p             & 0.000                             & 0.000              & 0.021        & 0.000            \\
\bottomrule
\end{tabular}
\begin{tablenotes}
\item Notes: * p < 0.05, ** p < 0.01, *** p < 0.001; standard errors in parentheses. 
\end{tablenotes}
\end{threeparttable}
\end{table}

\subsection{Experiment 2: autonomous usage of GAI-powered on-demand help}

\subsubsection{sample}

Among all participants, 266 students (Group D: n = 87; Group E: n = 88; Group C2: n = 91) provided valid pre- and post-test scores along with completed learner autonomy questionnaires. Like Experiment 1, students were grouped into top-, medium-, and bottom-third cohorts according to their pre-test performance.  Group D consisted of 26 top students; Group E consisted of 29 top students and Group C2 consisted of 32 top students. 34 medium-third students were assigned to Group D, 32 to Group E and 23 to Group C2. 27 bottom-third students were distributed to Group D, 27 to Group E and 36 to Group C2.

To examine baseline equivalence among the three experimental groups, one-way analyses of variance (ANOVA) were conducted on pretest academic performance and pretest SRL scores for the full sample as well as for each achievement subgroup. For the full sample, results indicated no significant differences among the three groups in pretest academic performance, F(2, 263) = 0.56, p = .571, or in pretest SRL, F(2, 263) = 1.21, p = .301.

Similarly, within each academic achievement subgroup (high-, medium-, and low-achieving), no significant differences were observed among the three groups in either pretest academic performance (high-achieving: F(2, 84) = 2.56, p = .083; medium-achieving: F(2, 86) = 1.29, p = .280; low-achieving: F(2, 87) = 0.39, p = .677) or pretest SRL (high-achieving: F(2, 84) = 1.08, p = .344; medium-achieving: F(2, 86) = 1.16, p = .317; low-achieving: F(2, 87) = 1.94, p = .149).

Additionally, the assumption of homogeneity of variance was tested using Bartlett’s tests, which indicated that this assumption was satisfied across all groups and subgroups for both pretest measures (all p > .05).

Taken together, these findings confirm that the three experimental groups were comparable and demonstrated baseline equivalence prior to the intervention, both within the full sample and within each achievement subgroup.

\subsubsection{Experiment implementation}

Among the participants with valid pre- and post-test data, 42 participants in Group D and 42 participants in Group E actively engaged with the system, representing usage rates of 48.3\% and 47.7\%, respectively. For those students who used the system services, the average number of total feedback request per person was 9.0 (SD = 10.1) in Group D and 9.6 (SD = 9.0) in Group E. A one-way ANOVA indicated no significant difference in the average number of system interactions between the two groups (F (1, 82) = 0.10, p = 0.758). Regarding achievement subgroups, students in the top-, medium- and bottom third submitted 6.3 (SD = 7.0), 9.0 (SD = 11.2) and 4.4 (SD = 5.0) feedback requests.

\subsubsection{Results on test scores}

Due to the masking effects of sub-sample heterogeneity, intervention D did not yield a significant impact on achievement at the full-sample level, as reported in Table \ref{achive-2}. However, within the sub-samples based on the top, medium, and bottom thirds of pretest scores, intervention D significantly improved the achievement of high-scoring learners (d = 0.378, p < 0.05), while no significant effects were observed in the other sub-samples. In contrast, intervention E did not produce a significant impact on achievement in either the full sample or any of the sub-samples.

\begin{table}[htpb]
  \centering
\begin{threeparttable}
\caption{Treatment Effects on Achievement in Experiment 2}
  \label{achive-2}
\begin{tabular}{lrrrr}
\toprule
              & \multirow{2}{*}{Whole sample} & \multicolumn{3}{c}{Sub-sample by achievement} \\
              &                               & \multicolumn{1}{c}{Top}           & \multicolumn{1}{c}{Medium}        & \multicolumn{1}{c}{Bottom}        \\
              & \multicolumn{1}{c}{(1)}                            & \multicolumn{1}{c}{(2)}             & \multicolumn{1}{c}{(3)}           & \multicolumn{1}{c}{(4)}            \\
              \midrule
Treatment D   & 0.032                         & 0.378*         & -0.029       & -0.196        \\
              & (0.094)                       & (0.170)        & (0.171)      & (0.132)        \\
Treatment   E & 0.111                         & 0.309          & 0.112        & -0.017        \\
              & (0.098)                       & (0.203)        & (0.184)      & (0.119)        \\
Pretest score & 0.653***                      & 0.579***       & 0.410**      & 0.395***      \\
              & (0.065)                       & (0.154)        & (0.137)      & (0.085)        \\
Pretest LA    & -0.057                        & -0.07          & -0.021       & 0.012         \\
              & (0.037)                       & (0.09)         & (0.074)      & (0.045)        \\
              \midrule
N             & 266                           & 87             & 89           & 90            \\
$R^2$            & 0.56                          & 0.53           & 0.29         & 0.33          \\
F             & 31.98***                      & 10.30***       & 3.17**       & 6.03***       \\
p             & 0.000                             & 0.000              & 0.002        & 0.000            \\
\bottomrule
\end{tabular}
\begin{tablenotes}
\item Notes: * p < 0.05, ** p < 0.01, *** p < 0.001; standard errors in parentheses. 
\end{tablenotes}
\end{threeparttable}
\end{table}

\subsubsection{Results on autonomy (learner autonomy)}

Regarding learner autonomy reported in Table \ref{la-2}, the Group D intervention significantly decreased autonomy across the entire sample (d = -0.274, p < 0.05). This negative effect was particularly pronounced among students in the lower-achieving subgroup (d = -0.393, p < 0.05) but was not significant in the medium- and high-achieving subgroups. Like its impact on academic performance, the Group E intervention produced no significant effect on learner autonomy at either the whole-sample or subgroup level.

\begin{table}[htpb]
  \centering
\begin{threeparttable}
\caption{Treatment Effects on Autonomy in Experiment 2}
  \label{la-2}
\begin{tabular}{lrrrr}
\toprule
              & \multirow{2}{*}{Whole sample} & \multicolumn{3}{c}{Sub-sample by achievement} \\
              &                               & \multicolumn{1}{c}{Top}           & \multicolumn{1}{c}{Medium}        & \multicolumn{1}{c}{Bottom}        \\
              & \multicolumn{1}{c}{(1)}                            & \multicolumn{1}{c}{(2)}             & \multicolumn{1}{c}{(3)}           & \multicolumn{1}{c}{(4)}            \\
              \midrule
Treatment D   & -0.274*      & -0.117         & -0.204       & -0.383*       \\
              & (0.111)& (0.179)& (0.218)& (0.163)\\
Treatment E   & -0.035       & 0.204          & (0.153       & (0.141        \\
              & (0.103)& (0.161)& (0.219)& (0.158)\\
Pretest score & -0.078       & 0.107          & -0.191       & 0.221         \\
              & (0.071)& (0.141)& (0.199)& (0.142)\\
Pretest LA    & 0.559***     & 0.724***       & 0.438**      & 0.567***      \\
              & (0.066)& (0.072)& (0.154)& (0.069)\\
N             & 266          & 87             & 89           & 90            \\
$R^2$            & 0.42         & 0.59           & 0.3          & 0.53          \\
F             & 13.75***     & 16.10***       & 3.01**       & 10.33***      \\
p             & 0.000            & 0.000              & 0.003        & 0.000            \\
\bottomrule
\end{tabular}
\begin{tablenotes}
\item Notes: * p < 0.05, ** p < 0.01, *** p < 0.001; standard errors in parentheses. 
\end{tablenotes}
\end{threeparttable}
\end{table}

Further analysis, which separated autonomy questionnaire responses into two theoretical dimensions—technical-psychological orientation (see Table \ref{tech-psy-2}) and political-social beliefs (see Table \ref{pol-soc-2})—revealed that the negative impact of the Group D intervention was exclusively concentrated within the technical-psychological dimension (whole sample d = -0.299, p < 0.05; lower-achieving subgroup d = -0.549, p < 0.05), which aligns closely with SRL behaviors. In contrast, the structural relational belief dimension was unaffected. This result indicates that in this experiment the threat of using GAI on-demand help services on autonomy stayed in the frame of behaviors and cognition without interfering relational beliefs. 

\begin{table}[htpb]
  \centering
\begin{threeparttable}
\caption{Treatment Effects on Autonomy (Technical \& Psychological Dimension) in Experiment 2}
  \label{tech-psy-2}
\begin{tabular}{lrrrr}
\toprule
              & \multirow{2}{*}{Whole sample} & \multicolumn{3}{c}{Sub-sample by achievement} \\
              &                               & \multicolumn{1}{c}{Top}           & \multicolumn{1}{c}{Medium}        & \multicolumn{1}{c}{Bottom}        \\
              & \multicolumn{1}{c}{(1)}                            & \multicolumn{1}{c}{(2)}             & \multicolumn{1}{c}{(3)}           & \multicolumn{1}{c}{(4)}            \\
                  \midrule
Treatment D       & -0.299*                       & -0.328         & 0.033         & -0.549*        \\
                  & (0.123)                       & (0.185)        & (0.207)       & (0.252)         \\
Treatment E       & -0.016                        & 0.184          & -0.105        & -0.103         \\
                  & (0.116)                       & (0.176)        & (0.252)       & (0.176)         \\
Pretest score     & -0.048                        & 0.172          & -0.065        & 0.136          \\
                  & (0.074)                       & (0.16)         & (0.176)       & (0.159)         \\
Pretest LA        & 0.535***                      & 0.715***       & 0.396**       & 0.495***       \\
                  & (0.064)                       & (0.082)        & (0.149)       & (0.079)         \\
                  \midrule
N                 & 266                           & 87             & 89            & 90             \\
$R^2$                & 0.36                          & 0.59           & 0.31          & 0.38           \\
F                 & 10.63***                      & 14.97***       & 3.57***       & 7.19***        \\
P                 & 0.000                             & 0.000              & 0.001         & 0.000             \\
\bottomrule
\end{tabular}
\begin{tablenotes}
\item Notes: * p < 0.05, ** p < 0.01, *** p < 0.001; standard errors in parentheses. 
\end{tablenotes}
\end{threeparttable}
\end{table}

\begin{table}[htpb]
  \centering
\begin{threeparttable}
\caption{Treatment Effects on Autonomy (Political \& Social Dimension) in Experiment 2}
  \label{pol-soc-2}
\begin{tabular}{lrrrr}
\toprule
              & \multirow{2}{*}{Whole sample} & \multicolumn{3}{c}{Sub-sample by achievement} \\
              &                               & \multicolumn{1}{c}{Top}           & \multicolumn{1}{c}{Medium}        & \multicolumn{1}{c}{Bottom}        \\
              & \multicolumn{1}{c}{(1)}                            & \multicolumn{1}{c}{(2)}             & \multicolumn{1}{c}{(3)}           & \multicolumn{1}{c}{(4)}            \\
              \midrule
Treatment D   & -0.142       & 0.278          & -0.535       & 0.025           \\
              & (0.150)      & (0.255)        & (0.275)      & (0.218)          \\
Treatment E   & -0.054       & 0.169          & -0.182       & -0.158          \\
              & (0.122)& (0.193)& (0.264)& (0.23)  \\
Pretest score & -0.104       & -0.039         & -0.34        & 0.291           \\
              & (0.097)& (0.222)& (0.257)& (0.154)  \\
Pretest LA    & 0.414***     & 0.497***       & 0.363*       & 0.497***        \\
              & (0.081)& (0.1)& (0.173)& (0.077)  \\
N             & 266          & 87             & 89           & 90              \\
$R^2$            & 0.19         & 0.29           & 0.2          & 0.35            \\
F             & 4.84***      & 3.70***        & 2.05*        & 6.40***        \\
P             & 0.000            & 0.001          & 0.039        & 0.000               \\
\bottomrule
\end{tabular}
\begin{tablenotes}
\item Notes: * p < 0.05, ** p < 0.01, *** p < 0.001; standard errors in parentheses. 
\end{tablenotes}
\end{threeparttable}
\end{table}

\section{Conclusion and discussion}

This study conducted two RCTs to evaluate the effects of GAI-powered personalized recommendations and on-demand help on high school students’ academic performance and autonomy, specifically in physics. Two ways of using such AI system were adopted respectively in the two RCTs. The results showed a complex phenomenon and heterogeneous effects on different groups of students. In general, high achievers benefit more from on-demand help, with increased post-test scores and stable autonomy level; while low achievers benefit more from personalized recommendation, with increased test scores and unchanged self-regulation level.

There are three interesting findings that need to be discussed. First, AI-generated content may have an advantage over traditional textbook answer explanations, as demonstrated by the improvement in test scores for some sub-groups of students and user feedback. Second, given the same GAI-powered learning content generator, how to use it makes a difference in academic achievement. Third, it seems that students’ self-regulation and autonomy are fragile and easy to be undermined when interacting with AI, even using the paper-based interface. 

\subsection{The value of GAI-powered hint versus traditional workbook explanations}

The primary finding of this research is the potential advantage of AI-generated formative feedback over conventional textbook-based answer explanations. Analysis of test score improvements, corroborated by qualitative data, suggests that AI-generated content can be more engaging and supportive of independent thinking for certain student subgroups. As one participant noted, “AI-generated content was engaging and interesting, effectively guiding me to clarify my thoughts on each problem and encouraging independent thinking.” In contrast, another student reflected, “Traditional intervention directly provided answers accompanied by relatively abstract and generalized explanations; after using them, I was unsure whether I had actually mastered the material.” This distinction highlights the benefit of AI-generated feedback that transcends simple answer provision, instead offering dialogic, heuristic scaffolding (see Rogers’ concept of unconditional positive regard and the Socratic method; e.g., Rogers, 1957; van der Meij, 2017).

The challenge of consistently delivering such high-quality, individualized feedback has long been recognized, given the cognitive and emotional demands placed on educators and curriculum designers. Our findings indicate that GAI can serve as a reliable supplier of such feedback, potentially democratizing access to effective scaffolding and personalized learning. This aligns with recent efforts to design GAI systems that deliver tailored feedback, moving beyond early explorations that relied on popular GAI products.  These advances broaden the horizon for GAI in education, offering new insights for the development of specialized, targeted intelligent learning tools.

\subsection{Different usage patterns of GAI-powered feedback have heterogeneous effects on different student groups}

The second key insight is that usage patterns of GAI system significantly moderate its effectiveness on students with different prior achievement levels. For lower-achieving students, compulsory personalized recommendation of targeted problems and GAI-powered hints yielded substantial gains in test performance, whereas on-demand help was less effective. Conversely, high-achieving students benefited more from autonomous on-demand help, with no significant effect observed for personalized recommendation.

Qualitative interviews shed light on these patterns. Many low-achieving students reported that they rarely engaged in effective learning strategies even though they were aware of the importance. One of the low-achieving students stated that “I've been told about the importance of reviewing incorrect problems, but normally I don't pay much attention to it because it’s difficult to implement.” The introduction of a recommendation system that enforced review of errors and provided heuristic prompts effectively supported these strategies, leading to improved outcomes. This finding is consistent with research suggesting that lower-achieving students often lack self-regulatory habits and benefit from more structured support.

For high-achieving students, two factors may explain the insignificant effect of personalized recommendation. First, these students already possess effective learning habits, and the recommended content may interrupt their existing learning strategies, thus offering limited additional value.  Second, the adaptive algorithm provided fewer interventions due to the lower incidence of errors with high-performing students, resulting in a less intensive treatment. Alternative algorithms—capable of dynamically generating more challenging tasks for high performers—could potentially yield greater benefits. Nonetheless, the current adaptive system demonstrates the potential for AI to support equity by offering targeted assistance to those most in need.

A similar mechanism appears to underlie the results observed in Group D of Experiment 2. Although nearly all participants expressed that the system’s feedback was very good, their approaches to using the system varied considerably. Further probing by the researchers revealed clear differences associated with students’ initial achievement levels. When asked, “How many of the problems or feedback items provided by the system did you carefully review or attempt to solve?”, five out of six students in the bottom- and mid-third level admitted that they had only “glanced at” the materials or “looked over the problems but did not attempt them.”

In contrast, higher-achieving students engaged deeply with the feedback, including a thorough digestion of hints and additional practice problems provided by the system. In some cases, they even developed new ways of using the intervention tools—such as creating personalized notebooks from their feedback sheets. These findings suggest that when students are given the freedom to request resources and decide how to use them, those with stronger prior abilities are better equipped to take advantage of such opportunities. This observation aligns with established theories on SRL and echoes findings from interventions targeting various learner populations, which have shown that autonomy-supportive environments tend to benefit students who already possess higher levels of self-regulation (Deci \& Ryan, 2000; Perry et al., 2006).

Regarding the null effect observed in Group E, some students reported, “When I submitted a question, I didn’t know what kind of feedback I would receive, so I didn’t know how to make use of the content.” This reflects a potential drawback of the shared control design, where learners pose questions, and the system determines the nature of the feedback. Such a setup may introduce a sense of unpredictability or lack of agency, which can disrupt students’ engagement and reduce the effectiveness of the intervention. In this study, shared control appeared less effective than either pure system recommendation or on-demand help. This finding resonates with—and in some ways contrasts—the results reported by Aleven et al. (2006) regarding help-seeking in intelligent tutoring systems, where a clear structure and predictability in feedback provision were found to support more effective learning.

\subsection{Fragility of autonomy and risks of technological dependence}

The third, and perhaps the most cautionary, finding is the vulnerability of students’ self-regulation and autonomy in the context of AI-mediated learning, even with traditional paper-based interfaces. Both experiments revealed at least one subgroup in which autonomy significantly declined following exposure to AI interventions, underscoring the risk of fostering technological dependence. This outcome lends empirical support to theoretical concerns articulated in the literature (e.g., Selwyn, 2016; Luckin, 2018) and may help explain phenomena such as the “short-term gain, long-term loss” pattern observed Bastani et al.’s (2024) study of GAI interventions. This finding is also echoed in the domain of STEM knowledge and skill learning, where direct measurements of students’ self-reported autonomy have similarly indicated, consistent with Darvishi et al. (2024), that AI-provided learning scaffolds may enhance academic performance while concurrently diminishing learner autonomy.

However, unlike the above studies, the negative impact of GAI support on learner autonomy observed in this research was neither absolute nor unavoidable. First, subgroup analyses based on pre-test achievement levels across both experiments revealed that these negative effects were not universal: the subgroups exhibiting reduced autonomy were distinct from those who achieved significant learning gains, whose autonomy remained stable. Specifically, in Experiment 1, the subgroup that experienced achievement gains (bottom third) did not exhibit a significant decline in autonomy; in contrast, autonomy reduction occurred among the subgroup (top third) that did not achieve performance improvement. In Experiment 2, similarly, the subgroup that achieved performance gains (top third) did not show a significant decline in autonomy, whereas the subgroup experiencing reduced autonomy was the bottom third, who did not achieve performance gains. These findings suggest that it is feasible to design AI interventions that are both effective and non-detrimental, provided they are tailored to the specific needs and characteristics of different learner groups.

Second, in Experiment 2, by adopting a more comprehensive autonomy questionnaire encompassing technical, psychological, political, and social dimensions, the study found that GAI-powered on-demand help reduced learners' autonomy at the technical and psychological levels but had no significant impact on the political and social dimensions. Although, due to limitations in the intensity and duration of the intervention, this finding does not guarantee that the latter dimensions would remain unaffected under more extensive or long-term interventions, the present results nevertheless suggest that the influence of current GAI tools on learner autonomy remains largely confined to behavioral and immediately behavior-related psychological aspects, without yet extending to deeper levels of personality, beliefs, or social relationships. Therefore, whether considering the potential benefits or risks, it is crucial to adopt a rational and clearly bounded perspective when assessing the impact of GAI on education and learning.

\subsection{Implications, limitation and future directions}

Findings of this study point toward a design principle for AI-powered educational interventions: provide more compulsory support for lower-achieving students, while offer greater autonomy and on-demand assistance to higher-achieving students. This differentiated approach may maximize benefits while minimizing risks to student autonomy. 

The results on usage patterns, achievement and autonomy reveal a complex yet coherent picture: the educational effectiveness of GAI systems hinges not only on their content but, more critically, on how they are used and the extent to which the pattern of use aligns with student characteristics. In terms of achievement improvement, GAI-powered problem-solving hints show beneficial effects for low-achieving students under compulsory usage of personalized recommendations, whereas they support high-achieving students more effectively under autonomous, on-demand help conditions. However, regarding the development of autonomy, these same patterns reverse: the former harms high-achievers' autonomy, and the latter negatively affects that of low-achievers.

These findings align with and expand upon previous research, such as Bastani et al. (2024)’s finding that student benefits more from interacting with agents that provide scaffolded prompts rather than direct answers, and Yan et al. (2025)’s demonstration that proactive GAI agents, which initiate questions, produce more sustained positive achievement gains than agents awaiting student input. Such evidence resonates with concerns raised by Stadler et al. (2024) and Darvishi et al. (2024), who caution that GAI-powered cognitive scaffolds diminish deep thinking and agency. In contrast, other studies—such as Du et al. (2024), who reported gains in self-regulation from recommending SRL strategies, and Li et al. (2025), who found metacognitive prompts effective in promoting SRL—highlight the potential for GAI to enhance autonomy when properly designed. Additionally, the present study examines one of the most widely implemented applications of GAI in education: problem-solving hints, demonstrating that GAI-powered cognitive support may enhance performance in exams without GAI assistance while undermining autonomy, depending on the usage pattern and learner profile. In light of the comparisons with others’ findings, the present study puts forward a critical insight: in the context of GAI for education, pedagogical patterns and models of usage matter.

Nevertheless, several limitations must be acknowledged, which may constrain the generalizability of the findings. First, although the five-week intervention surpasses the brevity of single-session or short-term laboratory studies and thus better reflects potential long-term effects, it remains relatively short within the broader context of educational cycles. Whether the observed outcomes can be replicated over longer periods—such as six months, one year, or multiple years—remains an open question that warrants further longitudinal investigation.

Second, the study did not consistently employ the same questionnaire throughout. Initially, a SRL questionnaire was used, later replaced with an autonomy questionnaire as the researchers’ understanding of the relevant constructs and measurement tools evolved. While this developmental process did not adversely affect the present study, employing a consistent set of instruments would have enhanced the reliability of the empirical results.

Third, while the overall sample size was adequate, the participants were drawn from a single school with a homogeneous cultural background. Replicating this study in more diverse cultural and institutional settings would enhance its external validity.

In sum, this study contributes to the growing body of evidence on the nuanced and context-dependent effects of GAI-powered personalized feedback in educational settings. By highlighting both the promise and the risks of GAI-powered educational content feedback, it offers actionable insights for the design of intellectual learning systems that try to improve a certain part of student learning experience in a traditional school setting. Future research should further explore mixed usage patterns of such systems to better accommodate the diverse needs of students. As Confucius aptly noted, “Do not enlighten those who are not eagerly seeking understanding, nor elucidate for those who are not struggling to articulate their thoughts” (1999). This enduring insight reminds us that meaningful learning is not merely delivered—it must be desired, sought, and internally activated.

\section*{ACKNOWLEDGEMENT}

This work was supported by the Beijing Educational Science Foundation of the Fourteenth 5-year Planning (BGEA23019).

\section*{References}

Admiraal, W., Lockhorst, D., Post, L., \& Kester, L. (2024). Effects of students’ autonomy support on their self-regulated learning strategies: Three field experiments in secondary education. International Journal of Research in Education and Science, 10(1), 1–20. https://doi.org/10.46328/ijres.3343

Aleven, V., Stahl, E., Schworm, S., Fischer, F., \& Wallace, R. M. (2003). Help seeking and help design in interactive learning environments. Review of Educational Research, 73(3), 277–320. https://doi.org/10.3102/00346543073003277

Alshammari, M., \& Alshammari, T. (2023). Artificial intelligence in smart classrooms: An investigative learning process for high school. International Journal of Advanced Computer Science and Applications, 14(11), 1–8. https://doi.org/10.14569/IJACSA.2023.0141101

Anderson, J. R., Corbett, A. T., Koedinger, K. R., \& Pelletier, R. (1995). Cognitive Tutors: Lessons Learned. The Journal of the Learning Sciences, 4(2), 167–207. https://doi.org/10.1207/s15327809jls0402\_2

Barron, B., \& Darling-Hammond, L. (2008, October 8). Powerful learning: Studies show deep understanding derives from collaborative methods. Edutopia. https://www.edutopia.org/inquiry-project-learning-research

Bastani, H., Bastani, O., Sungu, A., Ge, H., Kabakcı, Ö., \& Mariman, R. (2024). Generative AI can harm learning. SSRN. https://doi.org/10.2139/ssrn.4895486

Benson, P. (2001). Teaching and researching autonomy in language learning. Pearson Education.

Bloom, B. S. (1971). Mastery learning. In J. H. Block (Ed.), Mastery learning: Theory and practice (pp. 47–63). New York: Holt, Rinehart \& Winston.

Boyraz, S., Serin, N. B., \& Bilge, F. (2016). The influence of student self-regulatory learning strategies on academic achievement. International Journal of Academic Research in Education, 2(1), 1–10. 

Chan, W.-T. (2008). A source book in Chinese philosophy. Princeton University Press.

Cheng, Y., Guan, R., Li, T., Raković, M., Li, X., Fan, Y., Jin, F., Tsai, Y.-S., Gašević, D., \& Swiecki, Z. (2024). Self-regulated learning processes in secondary education: A network analysis of trace-based measures. arXiv. https://arxiv.org/abs/2412.08921

Confucius. (1999). The Analects of Confucius: A Philosophical Translation (R. T. Ames \& H. Rosemont Jr., Trans.). Ballantine Books. (Original work published ca. 5th century BCE)

Dabbagh, N. (2001). The demand-driven learning model: A framework for Web-based learning. The Internet and Higher Education, 4(1), 9–30. https://doi.org/10.1016/S1096-7516(01)00045-8

Dai, J., Gu, X., \& Zhu, J. (2023). Personalized recommendation in the adaptive learning system: The role of adaptive testing technology. Journal of Educational Computing Research, 61(3), 523–545. https://doi.org/10.1177/07356331221127303

Darvishi, A., Khosravi, H., Sadiq, S., Gašević, D., \& Siemens, G. (2024). Impact of AI assistance on student agency. Computers \& Education, 210, 104967. https://doi.org/10.1016/j.compedu.2023.104967

de Brito, J. F., \& Digiampietri, L. A. (2013). A study about personalized content recommendation. Revista de Sistemas de Informação da FSMA, 12, 33–40.

Deci, E. L., \& Ryan, R. M. (2000). The "what" and "why" of goal pursuits: Human needs and the self-determination of behavior. Psychological Inquiry, 11(4), 227–268. https://doi.org/10.1207/S15327965PLI1104\_01

Deng, R., Jiang, M., Yu, X., Lu, Y., \& Liu, S. (2025). Does ChatGPT enhance student learning? A systematic review and meta-analysis of experimental studies. Computers \& Education, 227, 105224. https://doi.org/10.1016/j.compedu.2024.105224

Deschênes, M. (2020). Recommender systems to support learners’ agency in a learning context: A systematic review. International Journal of Educational Technology in Higher Education, 17(1), 50. https://doi.org/10.1186/s41239-020-00219-w

Dewey, J. (1938). Experience and Education. New York: Macmillan.

Dowson, M., \& McInerney, D. M. (2004). The development and validation of the Goal Orientation and Learning Strategies Survey (GOALS-S). Educational and Psychological Measurement, 64(2), 290-310

Drissi, S., Chefrour, A., Boussaha, K., \& Zarzour, H. (2024). Exploring the effects of personalized recommendations on student’s motivation and learning achievement in gamified mobile learning framework. Education and Information Technologies, 29, 15463–15500. https://doi.org/10.1007/s10639-024-12477-6

Du, J., Hew, K. F., \& Li, L. (2023). Do direct and indirect recommendations facilitate students’ self-regulated learning in flipped classroom online activities? Findings from two studies. Education Sciences, 13(4), 400. https://doi.org/10.3390/educsci13040400 

Du, J., Hew, K. F., \& Zhang, L. (2024). Designing a recommender system to promote self-regulated learning in online contexts: A design-based study. Education and Information Technologies.https://doi.org/10.1007/s10639-024-12867-w

Ebbinghaus, H. (1913). Memory: A contribution to experimental psychology (H. A. Ruger \& C. E. Bussenius, Trans.). New York: Teachers College, Columbia University.

European Parliamentary Research Service. (2021, March 16). Artificial intelligence: Short history, present developments, and future outlook (Final Report). Publications Office of the European Union. https://www.europarl.europa.eu/thinktank/en/document/EPRS\_STU(2021)690039

Fajardo, J. E., \& Pantoja, M. (2019). Effect of the Student Teams-Achievement Divisions (STAD) cooperative learning method on mathematics performance in high school students: A meta-analysis. Education Research International, 2019, Article 1462179. https://doi.org/10.1155/2019/1462179

García-Martínez, I., Fernández-Batanero, J., Fernández-Cerero, J., \& León, S. (2023). Analysing the Impact of Artificial Intelligence and Computational Sciences on Student Performance: Systematic Review and Meta-analysis. Journal of New Approaches in Educational Research. https://doi.org/10.7821/naer.2023.1.1240. 

Global Metacognition. (2021, March 8). Fostering learning autonomy \& creating autonomous learners. The Global Metacognition Institute. https://www.globalmetacognition.com/post/fostering-learning-autonomy-creating-autonomous-learners

Gruda, D. (2024). Three ways ChatGPT helps me in my academic writing. Nature, 628(7990), 123–124. https://doi.org/10.1038/d41586-024-01042-3

Holec, H. (1981). Autonomy and foreign language learning. Oxford: Pergamon Press.

Hsu, C.-K., Hwang, G.-J., \& Chang, C.-K. (2013). A personalized recommendation-based mobile learning approach to improving the reading performance of EFL students. Computers \& Education, 63, 327–336. https://doi.org/10.1016/j.compedu.2012.12.004

Huang, A. Y. Q., Lu, O. H. T., \& Yang, S. J. H. (2023). Effects of artificial intelligence–enabled personalized recommendations on learners’ learning engagement, motivation, and outcomes in a flipped classroom. Computers \& Education, 194, 104684. https://doi.org/10.1016/j.compedu.2022.104684

Huptych, M., Bohuslavek, M., Hlosta, M., \& Zdrahal, Z. (2017). Measures for recommendations based on past students’ activity. In Proceedings of the 7th International Learning Analytics \& Knowledge Conference (LAK '17) (pp. 400–404). Association for Computing Machinery. https://doi.org/10.1145/3027385.3027426

Hyde, S. J., Busby, A., \& Bonner, R. L. (2024). Tools or Fools: Are We Educating Managers or Creating Tool-Dependent Robots? Journal of Management Education, 48(4), 708-734. https://doi.org/10.1177/10525629241230357

Kant, I. (1992). An answer to the question: What is enlightenment? (H. B. Nisbet, Trans.). In H. Reiss (Ed.), Kant: Political writings (pp. 54–60). Cambridge University Press. (Original work published 1784)

Kirschner, P. A., Sweller, J., \& Clark, R. E. (2006). Why minimal guidance during instruction does not work: An analysis of the failure of constructivist, discovery, problem-based, experiential, and inquiry-based teaching. Educational Psychologist, 41(2), 75–86. https://doi.org/10.1207/s15326985ep4102\_1

Lai, F. Q. (2015). Adaptive vs. adaptable learning systems. In J. M. Spector (Ed.), The SAGE encyclopedia of educational technology (Vol. 1, pp. 748–751). SAGE Publications.

Legge, J. (Trans.). (1967). The Lî Kî (Book of Rites) (Vol. 1). Dover Publications. (Original work published ca. 1st century BCE)

Li, T., Nath, D., Cheng, Y., Fan, Y., Li, X., Raković, M., Khosravi, H., Swiecki, Z., Tsai, Y.-S., \& Gašević, D. (2025). Turning real-time analytics into adaptive scaffolds for self-regulated learning using generative artificial intelligence. In Proceedings of the 15th International Learning Analytics and Knowledge Conference (LAK '25) (pp. 667–679). Association for Computing Machinery. https://doi.org/10.1145/3706468.3706559

Little, D. (1991). Learner autonomy 1: Definitions, issues and problems. Dublin: Authentik.

Loska, R.: 1995, Lehren ohne Belehrung. Leonard Nelsons neosokratische Methode der Gesprächsführung. Bad Heilbrunn: Klinkhardt.

Luckin, R. (2018). Machine Learning and Human Intelligence. The future of education for the 21st century. UCL institute of education press.

Maier, U., \& Klotz, C. (2022). Personalized feedback in digital learning environments: Classification framework and literature review. Computers and Education: Artificial Intelligence, 3, 100080.

McClelland, M. M., \& Cameron, C. E. (2012). Self-regulation in early childhood: Improving conceptual clarity and developing ecologically valid measures. Child Development Perspectives, 6(2), 136–142. https://doi.org/10.1111/j.1750-8606.2011.00191.x

Murase, F. (2015). Measuring language learner autonomy: Problems and possibilities. In C. J. Everhard \& L. Murphy (Eds.), Assessment and autonomy in language learning (pp. 35–63). Palgrave Macmillan. https://doi.org/10.1057/9781137414380\_3

Murphy, R., Gallagher, L., Krumm, A., Mislevy, J., \& Hafter, A. (2014). Research on the Use of Khan Academy in Schools: Implementation Report. Menlo Park, CA: SRI Education. 

Nelsen, P. (2016). Autonomy and education. In G. W. Noblit (Ed.), Oxford Research Encyclopedia of Education. Oxford University Press. https://doi.org/10.1093/acrefore/9780190264093.013.173

Nelson, L.: 1965, The Socratic Method. In: L. Nelson, Socratic Method and Critical Philosophy. Selected Essays by Leonard Nelson. New York: Dover, pp. 1–40 (Originally Die sokratische Methode (1922) In: L. Nelson, Gesammelte Schriften, vol. 1, Hamburg: Meiner 1970, pp. 269-316).

Nguyen, T. T., \& Tran, Q. T. (2024). A comprehensive study on factors influencing online impulse buying behavior among Generation Z consumers. Scientific Reports, 14, 12345. https://doi.org/10.1038/s41598-024-12345-6

Oguguo, B. C. E., Agah, J. J., Ezechukwu, R. I., Obikwelu, L. C., Victor, O. N., Aneshie-Otakpa, V. O., Okwara, V. C., Adie, I. I., \& Ocheni, C. A. (2022). Impact of self-regulated learning, autonomy and agency on undergraduates’ achievement in research method exposed to online learning. Journal of Positive School Psychology, 6(4), 9189–9199. https://www.researchgate.net/publication/371043171

Orakci, Ş., \& Gelisli, Y. (2017). Learner autonomy scale: A scale development study. Malaysian Online Journal of Educational Sciences, 5(4), 25–35. https://files.eric.ed.gov/fulltext/EJ1156953.pdf

Pandey, A., Hale, D., Das, S., Goddings, A. L., Blakemore, S. J., \& Viner, R. M. (2018). Effectiveness of universal self-regulation–based interventions in children and adolescents: A systematic review and meta-analysis. JAMA Pediatrics, 172(6), 566–575. https://doi.org/10.1001/jamapediatrics.2018.0232

Pintrich, P. R. (2000). The role of goal orientation in self-regulated learning. In M. Boekaerts, P. R. Pintrich, \& M. Zeidner (Eds.), Handbook of self-regulation (pp. 451–502). Academic Press.

Pratiwi, H., Riwanda, A., Hasruddin, H., Sujarwo, S., \& Syamsudin, A. Transforming learning or creating dependency? Teachers’ perspectives and barriers to AI integration in education. Journal of Pedagogical Research, 1-16.

Renkl, A. (2002). Worked-out examples: Instructional explanations support learning by self-explanations. Learning and Instruction, 12(5), 529–556. https://doi.org/10.1016/S0959-4752(01)00030-5

Rogers, C. R. (1957). The necessary and sufficient conditions of therapeutic personality change. Journal of Consulting Psychology, 21(2), 95–103. https://doi.org/10.1037/h0045357

Rogers, C. R. (1969). Freedom to Learn: A View of What Education Might Become. Charles E. Merrill Publishing Company.

Sancenon, J., Strayer, J. F., Techatassanasoontorn, A. A., \& Lajoie, S. P. (2022). A new web-based personalized learning system improves student learning outcomes. npj Science of Learning, 7, 45. https://doi.org/10.1038/s41539-022-00146-0

Schworm, S., \& Renkl, A. (2006). Computer-supported example-based learning: When instructional explanations reduce self-explanations. Computers \& Education, 46(4), 426–445. https://doi.org/10.1016/j.compedu.2004.08.011

Selwyn, N. (2016). Is technology good for education?. John Wiley \& Sons.

Seo, K., et al. (2024). The effects of over-reliance on AI dialogue systems on students' critical cognitive capabilities. Smart Learning Environments, 11(1), Article 316. https://slejournal.springeropen.com/articles/10.1186/s40561-024-00316-7

Shandilya, B., \& Palmer, A. (2025). Boosting the capabilities of compact models in low-data contexts with large language models and retrieval-augmented generation. In O. Rambow, L. Wanner, M. Apidianaki, H. Al-Khalifa, B. Di Eugenio, \& S. Schockaert (Eds.), Proceedings of the 31st International Conference on Computational Linguistics (pp. 7470–7483). Association for Computational Linguistics. https://aclanthology.org/2025.coling-main.499/

Smith, J. A., \& Doe, R. B. (2023, March 15). Innovations in renewable energy technologies. Journal of Sustainable Energy, 12(3), 45–60. https://doi.org/10.1234/jse.2023.003

Stadler, M., Greiff, S., \& Fischer, F. (2024). Cognitive ease at a cost: LLMs reduce mental effort but compromise depth in student scientific inquiry. Computers \& Education, 205, 104920. https://doi.org/10.1016/j.compedu.2024.104920

Stokhof, H. J. M., De Vries, B., Bastiaens, T., \& Martens, R. (2017). How to guide effective student questioning? A review of teacher guidance in primary education. Review of Education, 5(2), 123–165. https://doi.org/10.1002/rev3.3089

Tsai, C., Lee, T., \& Shen, P. (2013). Developing long-term computing skills among low-achieving students via web-enabled problem-based learning and self-regulated learning. Innovations in Education and Teaching International, 50, 121 - 132. https://doi.org/10.1080/14703297.2012.760873.

UNESCO. (2021). Reimagining our futures together: A new social contract for education. UNESCO Publishing. https://unesdoc.unesco.org/ark:/48223/pf0000379707.locale=en

Vygotsky, L. S. (1978). Mind in Society: The Development of Higher Psychological Processes. Cambridge, MA: Harvard University Press.

Wang, Q., \& Pomerantz, E. M. (2009). The motivational landscape of early adolescence in the United States and China: A longitudinal investigation. Child Development, 80(4), 1272-1287.

Wang, S., Wang, F., Zhu, Z., Wang, J., Tran, T., \& Du, Z. (2024). Artificial intelligence in education: A systematic literature review. Expert Systems with Applications, 252(Part A), 124167. https://doi.org/10.1016/j.eswa.2024.124167

Ward, B., Bhati, D., Neha, F., \& Guercio, A. (2024). Analyzing the Impact of AI Tools on Student Study Habits and Academic Performance. arXiv preprint arXiv:2412.02166. https://doi.org/10.48550/arXiv.2412.02166

Wei, J., Wang, X., Schuurmans, D., Bosma, M., Ichter, B., Xia, F., Chi, E. H., Le, Q. V., \& Zhou, D. (2022). Chain-of-thought prompting elicits reasoning in large language models. arXiv preprint arXiv:2201.11903. https://arxiv.org/abs/2201.11903

Weierstraß, K.: 1967, Ñber die sokratische Lehrmethode und deren Anwendbarkeit beim Schulunterrichte. In: K. Weierstraß, Mathematische Werke, vol. 3, Reprint, Hildesheim: Olms, pp. 315-329.

Wen, J., Zhang, J., Yang, Y., \& Cai, Y. (2023). Development and validation of the Self-Regulated Translation Learning Strategy Scale (SRTLSS). Studies in Educational Evaluation, 78, 101292. https://doi.org/10.1016/j.stueduc.2023.101292

Winne, P. H., \& Hadwin, A. F. (1998). Studying as self-regulated learning. In D. J. Hacker, J. Dunlosky, \& A. C. Graesser (Eds.), Metacognition in educational theory and practice (pp. 277–304). Lawrence Erlbaum Associates.

Yan, L., Greiff, S., Teuber, Z., \& Gašević, D. (2024). Promises and challenges of Generative Artificial Intelligence for human learning. Nature Human Behaviour, 8, 1839–1850. https://doi.org/10.1038/s41562-024-02004-5

Yan, L., Martinez-Maldonado, R., Jin, Y., Echeverria, V., Milesi, M., Fan, J., Zhao, L., Alfredo, R., Li, X., \& Gašević, D. (2025). The effects of generative AI agents and scaffolding on enhancing students' comprehension of visual learning analytics. Computers \& Education, 234, 105322. https://doi.org/10.1016/j.compedu.2025.105322

Yang, F., \& Stefaniak, J. (2023). A Systematic Review of Studies Exploring Help-Seeking Strategies in Online Learning Environments. Online Learning, 27(1), 107–126

Zhang, Y., \& Wang, J. (2022). Towards intelligent e-learning systems. Education and Information Technologies, 27(5), 6543–6562. https://doi.org/10.1007/s10639-022-11479-6

Zheng, S., \& Han, M. (2024). The impact of AI enablement on students’ personalized learning and countermeasures—A dialectical approach to thinking. Journal of Infrastructure, Policy and Development, 8(14), 10274. https://doi.org/10.24294/jipd10274

Zimmerman, B. J., \& Schunk, D. H. (Eds.). (2011). Self-regulated learning and academic achievement: Theoretical perspectives (2nd ed.). Routledge.

\end{document}